\documentclass[sigplan]{acmart}

\acmConference[PLDI'19]{ACM SIGPLAN Conference on Programming Languages}{June 22--28, 2019}{Phoenix, AZ, USA}
\acmYear{2019}
\acmISBN{} 
\acmDOI{} 
\startPage{1}

\setcopyright{none}

\usepackage{booktabs} 
\usepackage[ruled]{algorithm2e} 
\usepackage[utf8]{inputenc}

\SetAlFnt{\small}
\SetAlCapFnt{\small}
\SetAlCapNameFnt{\small}
\SetAlCapHSkip{0pt}
\IncMargin{-\parindent}

\usepackage{soul}
\usepackage{setspace}

\usepackage{subfigure}
\usepackage{listings}
\usepackage{siunitx}
\usepackage{url}
\usepackage{graphicx}

\usepackage{etex}

\usepackage{mathrsfs}
\usepackage{microtype}
\usepackage{numprint}
\usepackage{multirow}
\usepackage{arydshln}
\usepackage{lipsum}
\usepackage{pifont}
\usepackage{cleveref}

\usepackage{collab}

\definecolor{string}{rgb}{0.7,0.0,0.0}
\definecolor{comment}{rgb}{0.13,0.54,0.13}
\definecolor{keyword}{rgb}{0.0,0.0,1.0}

\definecolor{mygreen}{rgb}{0,0.6,0}
\definecolor{mygray}{rgb}{0.5,0.5,0.5}
\definecolor{mymauve}{rgb}{0.58,0,0.82}

\collabAuthor{dl}{magenta}{David Leopoldseder}
\collabAuthor{ap}{orange}{Aleksandar Prokopec}
\collabAuthor{ar}{blue}{Andrea Rosa}
\collabAuthor{gd}{red}{Gilles Duboscq}
\collabAuthor{pt}{green}{Petr Tuma}
\collabAuthor{yz}{purple}{Yudi Zheng}
\collabAuthor{wb}{brown}{Walter Binder}
\collabAuthor{lb}{violet}{Lubomír Bulej}

\usepackage[T1]{fontenc}

\usepackage{flushend}

\makeatletter
\newcommand{\removelatexerror}{\let\@latex@error\@gobble}
\makeatother

\usepackage{multirow}
\usepackage{lipsum}

\newcommand\prog[1]{{\sf #1}}
\newcommand\bench[1]{{\textit{#1}}}
\newcommand\metric[1]{{\sf #1}}

\graphicspath{{figures/}}

\usepackage{array}
\newcolumntype{L}[1]{>{\raggedright\let\newline\\\arraybackslash\hspace{0pt}}m{#1}}
\newcolumntype{C}[1]{>{\centering\let\newline\\\arraybackslash\hspace{0pt}}m{#1}}
\newcolumntype{R}[1]{>{\raggedleft\let\newline\\\arraybackslash\hspace{0pt}}m{#1}}

\begingroup
  \catcode`\.=\active
  \gdef.{\normalperiod\allowbreak}%
\endgroup

\newcommand\aepath[1]{
  \bgroup
    \let\normalperiod=.
    \catcode`\.=\active
    \everyeof{\noexpand}
    \endlinechar=-1
    \ttfamily\scantokens{#1}
  \egroup}

\usepackage{newunicodechar}
\newunicodechar{§}{\monospaced}
\def\monospaced#1§{\aepath{#1}}
\newunicodechar{¡}{\italicized}
\def\italicized#1¡{\textit{#1}}

\usepackage{lipsum}
\usepackage{wrapfig}

\usepackage{tikz}

\usetikzlibrary{calc}
\usetikzlibrary{patterns}

\usepackage{pgfplots}

\pgfplotsset{
    /pgfplots/ybar legend/.style={
    /pgfplots/legend image code/.code={%
       \draw[##1,/tikz/.cd,yshift=-0.25em]
        (0cm,0cm) rectangle (6pt,0.7em);},
   },
}

\pgfplotsset{compat=1.3}

\usepackage{xcolor}

\usepackage[inline]{enumitem}

\let\oldbibitem\bibitem
\def\bibitem{\vfill\oldbibitem}

\usepackage{afterpage}

\newcommand\forcameraready[1]{#1} 

\usepackage{balance}

\usepackage{xr}
\begin{document}

\title[On Evaluating the Renaissance Benchmark Suite]
{On Evaluating the Renaissance Benchmarking Suite:
Variety, Performance, and Complexity}


\author{Aleksandar Prokopec}
\orcid{0000-0003-0260-2729}             
\affiliation{
  \institution{Oracle Labs}            
  \country{Switzerland}                    
}
\email{aleksandar.prokopec@oracle.com}          

\author{Andrea Rosà}
\affiliation{
  \institution{Università della Svizzera italiana}           
  \country{Switzerland}                   
}
\email{andrea.rosa@usi.ch}         

\author{David Leopoldseder}
\affiliation{
  \institution{Johannes Kepler Universität Linz}           
  \country{Austria}                   
}
\email{david.leopoldseder@jku.at}         

\author{Gilles Duboscq}
\affiliation{
  \institution{Oracle Labs}           
  \country{Switzerland}                   
}
\email{gilles.m.duboscq@oracle.com}         

\author{Petr Tůma}
\affiliation{
  \institution{Charles University}           
  \country{Czech Republic}                   
}
\email{petr.tuma@d3s.mff.cuni.cz}         

\author{Martin Studener}
\affiliation{
  \institution{Johannes Kepler Universität Linz}           
  \country{Austria}                   
}
\email{martinstudener@gmail.com}         

\author{Lubomír Bulej}
\affiliation{
  \institution{Charles University}           
  \country{Czech Republic}                   
}
\email{bulej@d3s.mff.cuni.cz}         

\author{Yudi Zheng}
\affiliation{
  \institution{Oracle Labs}           
  \country{Switzerland}                   
}
\email{yudi.zheng@oracle.com}         

\author{Alex Villazón}
\affiliation{
  \institution{Universidad Privada Boliviana}           
  \country{Bolivia}                   
}
\email{avillazon@upb.edu}         

\author{Doug Simon}
\affiliation{
  \institution{Oracle Labs}           
  \country{Switzerland}                   
}
\email{doug.simon@oracle.com}         

\author{Thomas Wuerthinger}
\affiliation{
  \institution{Oracle Labs}           
  \country{Switzerland}                   
}
\email{thomas.wuerthinger@oracle.com}         

\author{Walter Binder}
\affiliation{
  \institution{Università della Svizzera italiana}           
  \country{Switzerland}                   
}
\email{walter.binder@usi.ch}         

\begin{abstract}
The recently proposed \emph{Renaissance} suite is composed of
modern, real-world, concurrent, and object-oriented workloads
that exercise various concurrency primitives of the JVM.
Renaissance was used to compare performance of two state-of-the-art,
production-quality JIT compilers (HotSpot C2 and Graal),
and to show that the performance differences are more significant
than on existing suites such as DaCapo and SPECjvm2008.

In this technical report, we give an overview of the experimental setup
that we used to assess the variety and complexity of the Renaissance suite,
as well as its amenability to new compiler optimizations.
We then present the obtained measurements in detail.
\end{abstract}

\ccsdesc[500]{Software and its engineering~General programming languages}
\ccsdesc[300]{Social and professional topics~History of programming languages}

\keywords{benchmarks, JIT compilation, parallelism}  

\maketitle



\section{Introduction}
\label{sec:app:intro}

\begin{figure*}
\includegraphics[width=\textwidth]{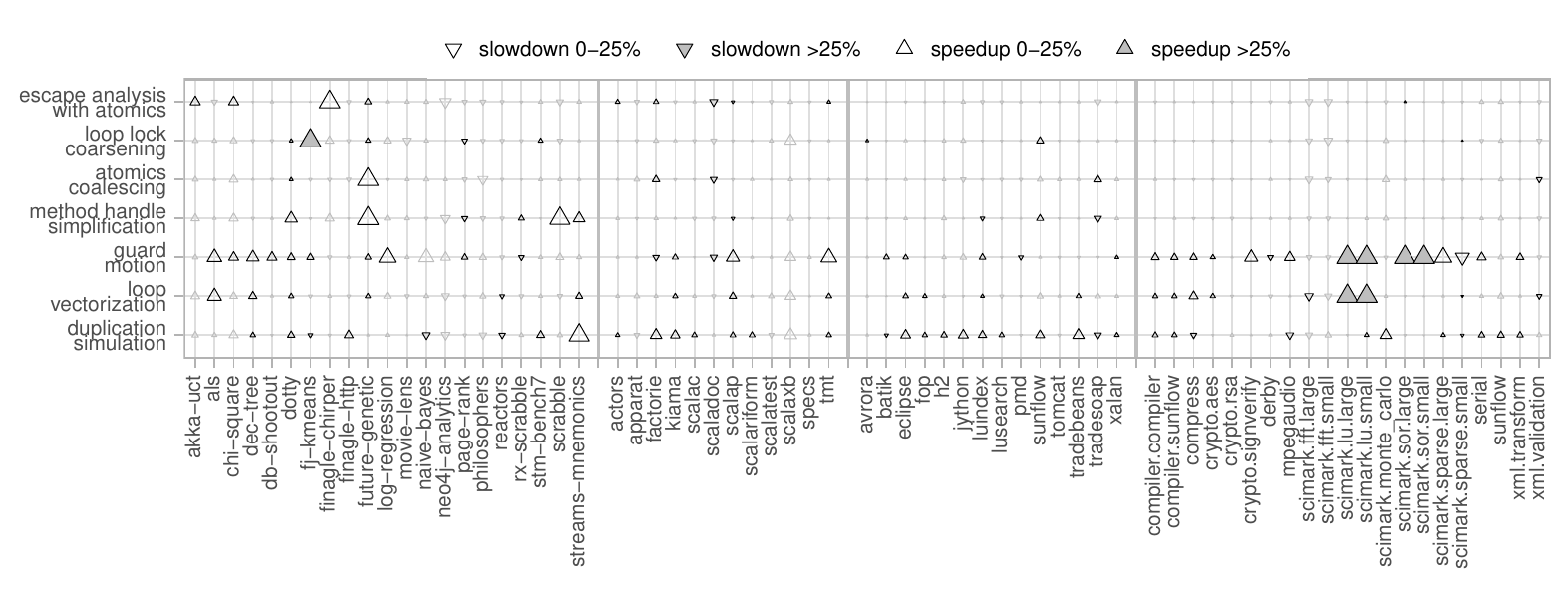}
\caption{
Optimization impact on individual benchmarks.
Results with black outline significant at $\alpha = 0.01$.
}
\label{fig:optimization-suite-impact}
\end{figure*}

The Renaissance suite~/cite{pldi-prokopec-19}
has proposed a set of new benchmarks for the JVM,
shown in Table \ref{tab:benchmarks},
which are focused on modern functional, concurrent and parallel applications and frameworks.
In related work, we evaluated Renaissance by determining a set of runtime metrics,
presented in Table \ref{tab:metrics},
which focus on traditional complexity indicators,
such as dynamic dispatch and object allocation rate,
as well as concurrency-focused behavior of the program.
Using a PCA analysis on these metrics, we showed that the benchmarks in the Renaissance suite
behave considerably different than other benchmark suites with respect to these metrics.
Furthermore, we have shown that some benchmarks in the new suite
indicate the need for new compiler optimizations.
Figure \ref{fig:optimization-suite-impact} shows the impact
of each of the seven optimizations that we studied,
across all the benchmarks from Renaissance,
as well as the existing DaCapo, Scalabench and SPECjvm2008 suites.
At the same time, we showed that Renaissance is comparable to these existing suites
in terms of its code complexity.

In this report, we give a more detail account of our experimental setup and our measurements.
We first explain the technical details of our experimental setup,
and we then present our experimental results, in terms of the metrics used in our PCA analysis,
performance comparison and the Chidamber \& Kemerer software complexity metrics.
We conclude the report by presenting some basic information
about the impact of the new optimizations on the warmup time of JIT-compiled code.


\begin{table*}
  \caption{Summary of benchmarks included in Renaissance.}
  \label{tab:benchmarks}
  \footnotesize
\begingroup
\setlength{\tabcolsep}{10pt} 
\renewcommand{\arraystretch}{1.2} 
  \begin{tabular}{
    l@{\hskip0.2cm} L{8.8cm} l l
  }
\toprule
   {\bf  Benchmark} & {\bf Description} & {\bf Focus}\\
\midrule
    \bench{akka-uct}          & Unbalanced Cobwebbed Tree computation using Akka \cite{akka}.
    & actors, message-passing \\\hdashline[0.5pt/1pt]
    \bench{als}               & Alternating Least Squares algorithm using Spark.
    & data-parallel, compute-bound \\\hdashline[0.5pt/1pt]
    \bench{chi-square}        & Computes a Chi-Square Test in parallel using Spark ML \cite{JMLR:v17:15-237}.
    & data-parallel, machine learning \\\hdashline[0.5pt/1pt]
    \bench{db-shootout}       & Parallel shootout test on Java in-memory databases.
    & query-processing, data structures \\\hdashline[0.5pt/1pt]
    \bench{dec-tree}          & Classification decision tree algorithm using Spark ML \cite{JMLR:v17:15-237}.
    & data-parallel, machine learning \\\hdashline[0.5pt/1pt]
    \bench{dotty}             & Compiles a Scala codebase using the Dotty compiler for Scala.
    & data-structures, synchronization \\\hdashline[0.5pt/1pt]
    \bench{finagle-chirper}   & Simulates a microblogging service using Twitter Finagle \cite{github-finagle}.
    & network stack, futures, atomics \\\hdashline[0.5pt/1pt]
    \bench{finagle-http}      & Simulates a high server load with Twitter Finagle \cite{github-finagle} and Netty \cite{github-netty}.
    & network stack, message-passing \\\hdashline[0.5pt/1pt]
    \bench{fj-kmeans}         & K-means algorithm using the Fork/Join framework \cite{Lea:2000:JFF:337449.337465}.                                       
    & task-parallel, concurrent data structures \\\hdashline[0.5pt/1pt]
    \bench{future-genetic}    & Genetic algorithm function optimization using Jenetics \cite{github-jenetics}.
    & task-parallel, contention \\\hdashline[0.5pt/1pt]
    \bench{log-regression}    & Performs the logistic regression algorithm on a large dataset. 
    & data-parallel, machine learning \\\hdashline[0.5pt/1pt]
    \bench{movie-lens}        & Recommender for the MovieLens dataset using Spark ML \cite{JMLR:v17:15-237}.
    & data-parallel, compute-bound \\\hdashline[0.5pt/1pt]
    \bench{naive-bayes}       & Multinomial Naive Bayes algorithm using Spark ML \cite{JMLR:v17:15-237}.
    & data-parallel, machine learning \\\hdashline[0.5pt/1pt]
    \bench{neo4j-analytics}   & Analytical queries and transactions on the Neo4J database \cite{github-neo4j}.
    & query processing, transactions \\\hdashline[0.5pt/1pt]
    \bench{page-rank}         & PageRank using the Apache Spark framework \cite{Zaharia:2010:SCC:1863103.1863113}.
    & data-parallel, atomics \\\hdashline[0.5pt/1pt]
    \bench{philosophers}      & Dining philosophers using the ScalaSTM framework \cite{scala-stm-expert-group}.
    & STM, atomics, guarded blocks \\\hdashline[0.5pt/1pt]
    \bench{reactors}          & A set of message-passing workloads encoded in the Reactors framework \cite{Prokopec:2015:ICE:2814228.2814245}.
    & actors, message-passing, critical sections \\\hdashline[0.5pt/1pt]
    \bench{rx-scrabble}       & Solves the Scrabble puzzle \cite{scrabble-benchmark} using the RxJava framework.
    & streaming \\\hdashline[0.5pt/1pt]
    \bench{scrabble}          & Solves the Scrabble puzzle \cite{scrabble-benchmark} using Java 8 Streams. 
    & data-parallel, memory-bound \\\hdashline[0.5pt/1pt]
    \bench{stm-bench7}        & STMBench7 workload \cite{Guerraoui:2007:SBS:1272996.1273029} using the ScalaSTM framework \cite{scala-stm-expert-group}.
    & STM, atomics \\\hdashline[0.5pt/1pt]
    \bench{streams-mnemonics} & Computes phone mnemonics \cite{state-of-scala} using Java 8 Streams.
    & data-parallel, memory-bound \\
\bottomrule
\end{tabular}
\endgroup
\end{table*}


\begin{table}
  \caption{Metrics considered during benchmark selection.}
  \label{tab:metrics}
  \footnotesize
\begingroup
  \begin{tabular}{
    l@{\hskip0.2cm} p{7cm}
  }
    \toprule
{\bf Name} & {\bf Description}\\
    \midrule
\metric{synch}  & §synchronized§ methods and blocks executed.\\
\hdashline[0.5pt/1pt]
\metric{wait} & Invocations of §Object.wait()§.\\
\hdashline[0.5pt/1pt]
\metric{notify} & Invocations of §Object.notify()§ and §Object.notifyAll()§ .\\
\hdashline[0.5pt/1pt]
\metric{atomic} & Atomic operations executed.\\
\hdashline[0.5pt/1pt]
\metric{park} & Park operations.\\
\hdashline[0.5pt/1pt]
\metric{cpu}  & Average CPU utilization (user and kernel).\\
\hdashline[0.5pt/1pt]
\metric{cachemiss}   &  Cache misses, including L1 cache (instruction and data), last-layer cache (LLC),
and translation lookaside buffer (TLB; instruction and data).\\
\hdashline[0.5pt/1pt]
\metric{object} & Objects allocated.\\
\hdashline[0.5pt/1pt]
\metric{array}  & Arrays allocated.\\
\hdashline[0.5pt/1pt]
\metric{method} & Methods invoked with §invokevirtual§, §invokeinterface§ or §invokedynamic§ bytecodes.\\
\hdashline[0.5pt/1pt]
\metric{idynamic}& §invokedynamic§ bytecodes executed.\\
    \bottomrule
\end{tabular}
\endgroup
\end{table}

\section{Analyzed Benchmarks}
\label{sec:app:benchmarks}


\begin{table}[t]
\caption{Benchmarks considered in the paper, including input size and number of operations executed (\# ops). }
\label{table:allworkloads}
\centering \footnotesize
\begin{tabular}{  p{2.2cm} p{1.1cm} | p{2.2cm} p{1.1cm} }
\hline
 {\bf Benchmark} & {\bf Input}  &  {\bf Benchmark} & {\bf Input}\\  
{\bf DaCapo~\cite{Blackburn:2006:DBJ:1167515.1167488}} & {\bf size}   &  {\bf ScalaBench~\cite{Sewe:2011:DCC:2048066.2048118}} & {\bf size}\\
\hline
\bench{avrora} & large  & \bench{actors} & huge\\
\hdashline[0.5pt/1pt]
\bench{batik} & large  & \bench{apparat} & gargantuan\\
\hdashline[0.5pt/1pt]
\bench{eclipse} & large  & \bench{factorie} & gargantuan\\
\hdashline[0.5pt/1pt]
\bench{fop} & default & \bench{kiama} & default\\  
\hdashline[0.5pt/1pt]
\bench{h2} & huge  & \bench{scalac} & large\\ 
\hdashline[0.5pt/1pt]
\bench{jython} & large  & \bench{scaladoc} & large\\
\hdashline[0.5pt/1pt]
\bench{luindex}  & default  & \bench{scalap} & large\\ 
\hdashline[0.5pt/1pt]
\bench{lusearch-fix} & large  & \bench{scalariform} & huge\\
\hdashline[0.5pt/1pt]
\bench{pmd} & large  & \bench{scalatest} & default\\
\hdashline[0.5pt/1pt]
\bench{sunflow} & large  & \bench{scalaxb} & huge\\ 
\hdashline[0.5pt/1pt]
\bench{tomcat} & huge  & \bench{specs} & large\\
\hdashline[0.5pt/1pt]
\bench{tradebeans} & huge  & \bench{tmt} & huge\\
\hdashline[0.5pt/1pt]
\bench{tradesoap} &  huge  &&\\
\hdashline[0.5pt/1pt]
\bench{xalan} & large   &&\\
\hline
\end{tabular}
\begin{tabular}{  p{2.2cm} >{\raggedleft\arraybackslash}p{1.1cm}  p{2.2cm} >{\raggedleft\arraybackslash}p{1.1cm} }
{\bf Benchmark} & \multirow{2}{*}{\bf \# ops} & \multirow{2}{*}{\bf Benchmark} & \multirow{2}{*}{\bf \# ops}\\
{\bf SPECjvm2008~\cite{specjvm2008}}  & &  &\\
\hline 
\bench{compiler.compiler} & \num{20}  & \bench{scimark.lu.small} & \num{125}\\
\hdashline[0.5pt/1pt]
\bench{compiler.sunflow}  & \num{20}& \bench{scimark.monte\_carlo} &\num{900}\\
\hdashline[0.5pt/1pt]
\bench{compress} & \num{50} & \bench{scimark.sor.large} & \num{15}\\
\hdashline[0.5pt/1pt]
\bench{crypto.aes} & \num{20} & \bench{scimark.sor.small} & \num{75}\\
\hdashline[0.5pt/1pt]
\bench{crypto.rsa} & \num{150}  & \bench{scimark.sparse.large} & \num{10}\\
\hdashline[0.5pt/1pt]
 \bench{crypto.signverify} & \num{125} & \bench{scimark.sparse.small} & \num{25}\\
 \hdashline[0.5pt/1pt]
\bench{derby} & \num{30} & \bench{serial} & \num{25}\\
\hdashline[0.5pt/1pt]
\bench{mpegaudio} & \num{50} & \bench{sunflow} & \num{30}\\
\hdashline[0.5pt/1pt]
\bench{scimark.fft.large}  & \num{10} & \bench{xml.transform} & \num{7}\\
\hdashline[0.5pt/1pt]
\bench{scimark.fft.small} & \num{100} & \bench{xml.validation} & \num{40}\\
\hdashline[0.5pt/1pt]
\bench{scimark.lu.large} & \num{4} \\
\hline
\end{tabular} 
\end{table}

Table~\ref{table:allworkloads} lists the benchmarks from the DaCapo, 
ScalaBench and SPECjvm2008 suites that were considered in the main paper,
along with the used input size (expressed as number of operations executed in SPECjvm2008).

\section{Experimental Setup for Metric Profiling and Principal Component Analysis}
\label{sec:app:pca:settings}


Here, we detail the experimental setup for the collection of metrics described in 
Table~\ref{tab:metrics} and analyzed in
Section~\ref{sec:pca} and~\ref{sec:compiler-optimizations} of the main paper.

The metrics are profiled
during a single steady-state benchmark execution. Before collecting the metrics,
we let the benchmarks warp-up until dynamic compilation
and GC ergonomics are stabilized, following the methodology 
of Lengauer et al.~\cite{Lengauer:2017:CJB:3030207.3030211}. 
We could not collect metrics for benchmarks \bench{tradebeans}, \bench{actors} and \bench{scimark.monte\_carlo}
either because bytecode instrumentation causes 
a premature workload termination with a §TimeoutException§ (\bench{tradebeans}, \bench{actors})
or because profiling takes an excessive amount of time, exceeding 7~days (\bench{scimark.monte\_carlo}).
Therefore we excluded such benchmarks from the PCA analysis
(Section~\ref{sec:pca} of the main paper).

We collect the metrics on a machine with two NUMA nodes,
each containing an Intel Xeon E5-2680 (2.7~GHz) processor
with 8~physical cores and 64~GB of RAM, 
running under Ubuntu~16.04.03~LTS (kernel GNU/Linux~4.4.0-112-generic \mbox{x86\_64}). 
We configure \prog{top} to sample CPU utilization only for the NUMA node
where the benchmark is executing, to increase the accuracy of the collected 
measurements (as the computational resources used by \prog{perf} and \prog{top} are not accounted). 
We disable Turbo Boost~\cite{turboboost} and Hyper-Threading~\cite{hyperThreading}. 
We use Java \mbox{OpenJDK~1.8.0\_161-b12}.

We collect the metrics in two runs,
profiling OS- and hardware-layer metrics
(\metric{cpu} and \metric{cachemiss})
in the first run on the original program,
and the other metrics in the second run
(using \prog{DiSL} instrumentation).
This way, we obtain more precise metrics at the OS- and hardware-layer,
which do not account for the execution of instrumentation code. 
During metric collection, no other CPU-, memory-, or IO-intensive application is executing
on the system to reduce measurement perturbations. 
In addition, we pin the execution to an exclusive NUMA node, 
to reduce performance interference caused by other running processes.

\section{Experimental Setup for Performance Evaluation}
\label{sec:appendix:charles}


Here, we describe the experimental setup for the performance evaluation described 
in Section~\ref{sec:evaluation} of the main paper.

The performance measurement experiments are conducted on 8-core Intel servers,
equipped with an Intel Xeon E5-2620v4 CPU (2.1 GHz, 8 cores, 20 MB cache, Hyper Threading disabled), 64 GB RAM, running Fedora Linux 27 (kernel 4.15.6).
For stable measurement, power management features are disabled and the processor is run at the nominal frequency.
Prior to each benchmark execution, the physical memory pool is randomized.
We use Oracle JDK 8u172 with Graal 1.0.0-rc9 as virtual machine.
The heap size is fixed at 12 GB with the G1 collector, and except for the selection of
individual compiler optimizations used to produce Figure~\ref{fig:optimization-suite-impact},
no other option is used.

For each benchmark and each optimization configuration, we execute the measurements in a new JVM 15 times. 
Each execution consists of a warm-up period of 5 minutes, followed by 60 seconds of steady-state execution, rounded up to the next complete benchmark iteration.
The duration of the warm-up period is chosen so that major performance fluctuations due to compilation happen before actual measurement (verified manually).
To provide for meaningful comparison across benchmarks, we always collect
the execution times of the main benchmark operation (we have modified
the SPECjvm2008 benchmark harness to achieve this, the benchmark
would normally report aggregated throughput).
Winsorized filtering is used to remove outliers from Figure~\ref{fig:optimization-suite-impact}.

\section{Collected Metrics}
\label{sec:app:metrics}


\forcameraready{
Table~\ref{tab:pcametricsext} reports the metrics
(listed in Table~\ref{tab:metrics}) collected on all analyzed benchmarks,
before being normalized by reference cycles. The experimental setup used for metric collection
is detailed in Section~\ref{sec:app:pca:settings} of the main paper.
}

\begin{table*}[ht]
\scriptsize
\begin{tabular}{l | r r r r r r r r r r r}
{\bf Benchmark}&{\bf synch} &{\bf wait} &{\bf notify}& {\bf atomic}& {\bf park}&{\bf cpu}& {\bf cachemiss}&{\bf object}&{\bf array}&{\bf method}&{\bf idynamic}\\
\hline
\multicolumn{12}{c}{\bf Renaissance}\\
\hdashline[0.5pt/1pt]
akka-uct&4.27E+05&1.00E+00&2.00E+00&1.18E+07&1.75E+05&94.45&6.24E+08&1.16E+08&6.00E+05&1.96E+09&0.00E+00\\
als&3.01E+06&1.15E+02&1.31E+04&1.81E+06&2.39E+03&58.90&9.63E+08&1.10E+08&2.38E+07&2.91E+09&0.00E+00\\
chi-square&1.52E+06&8.70E+01&4.30E+01&1.58E+05&7.20E+01&26.19&4.97E+08&1.73E+08&2.40E+07&2.39E+09&0.00E+00\\
db-shootout&7.28E+06&3.20E+01&0.00E+00&2.72E+07&4.01E+05&45.53&2.91E+09&2.16E+08&1.92E+08&1.11E+10&1.00E+06\\
dec-tree&5.83E+05&8.80E+01&1.37E+03&5.45E+05&5.35E+02&27.23&7.54E+08&2.50E+08&2.84E+07&2.96E+09&0.00E+00\\
dotty&5.63E+06&4.00E+00&2.56E+04&4.33E+04&0.00E+00&15.68&7.59E+08&4.92E+07&1.42E+07&1.26E+09&7.22E+06\\
finagle-chirper&1.29E+07&1.72E+03&1.74E+03&1.02E+08&1.72E+04&69.82&2.52E+09&1.43E+08&1.01E+07&4.44E+09&2.36E+03\\
finagle-http&2.72E+04&2.00E+01&0.00E+00&5.20E+04&6.66E+02&14.72&4.14E+08&2.81E+08&6.40E+04&3.09E+09&5.80E+02\\
fj-kmeans&1.01E+08&6.57E+02&6.62E+02&1.89E+04&1.19E+03&69.59&4.23E+08&1.35E+08&2.45E+03&7.08E+08&0.00E+00\\
future-genetic&6.72E+05&2.37E+04&2.40E+04&5.00E+07&1.59E+05&55.85&7.04E+08&2.11E+08&2.64E+05&1.58E+09&2.34E+06\\
log-regression&2.09E+05&1.09E+02&9.81E+02&4.77E+05&6.62E+02&24.83&6.58E+08&5.39E+07&1.68E+07&1.86E+09&0.00E+00\\
movie-lens&1.24E+07&5.78E+02&2.22E+05&3.11E+07&3.98E+04&44.17&3.37E+09&2.00E+08&2.58E+07&7.32E+09&2.16E+02\\
naive-bayes&2.33E+05&3.50E+01&1.09E+02&1.81E+04&1.32E+02&76.90&1.04E+09&3.60E+08&8.21E+07&3.65E+09&0.00E+00\\
neo4j-analytics&1.37E+07&4.23E+02&1.54E+05&2.05E+06&2.02E+02&59.74&1.05E+18&1.42E+09&3.29E+07&2.22E+10&2.49E+07\\
page-rank&2.63E+06&9.10E+01&1.26E+02&9.25E+06&1.38E+02&56.14&1.23E+09&2.01E+08&2.94E+06&5.15E+09&0.00E+00\\
philosophers&2.21E+06&1.52E+04&8.15E+04&1.18E+08&2.52E+04&99.21&1.16E+09&1.80E+08&4.81E+07&6.28E+09&0.00E+00\\
reactors&2.59E+08&0.00E+00&0.00E+00&1.76E+08&5.52E+06&56.62&4.22E+09&2.71E+08&1.86E+07&1.24E+10&0.00E+00\\
rx-scrabble&7.55E+06&0.00E+00&0.00E+00&7.49E+05&8.90E+01&25.10&8.11E+07&1.07E+07&0.00E+00&1.02E+08&1.71E+06\\
scrabble&2.00E+00&1.00E+00&1.00E+00&3.00E+01&1.00E+00&66.70&2.81E+08&5.65E+07&3.44E+06&4.99E+08&2.73E+07\\
stm-bench7&3.56E+03&1.00E+01&3.00E+00&2.92E+06&0.00E+00&49.44&3.48E+08&3.03E+07&2.96E+06&8.15E+08&0.00E+00\\
streams-mnemonics&0.00E+00&0.00E+00&0.00E+00&0.00E+00&0.00E+00&19.24&7.99E+08&2.04E+08&2.09E+08&1.15E+09&2.15E+07\\
\hline
\multicolumn{12}{c}{\bf DaCapo}\\
\hdashline[0.5pt/1pt]
avrora&7.22E+06&1.75E+06&1.68E+05&0.00E+00&0.00E+00&20.28&7.11E+17&7.22E+06&2.05E+06&2.78E+09&0.00E+00\\
batik&1.67E+06&6.00E+00&0.00E+00&7.00E+00&0.00E+00&24.54&1.96E+08&8.77E+05&2.03E+05&3.46E+07&0.00E+00\\
eclipse&6.80E+07&1.88E+04&3.60E+05&1.27E+05&0.00E+00&13.91&4.66E+09&9.06E+07&9.89E+07&2.41E+09&0.00E+00\\
fop&2.45E+06&0.00E+00&0.00E+00&1.60E+01&0.00E+00&6.25&5.10E+07&1.63E+06&7.12E+05&3.50E+07&0.00E+00\\
h2&7.76E+08&4.65E+03&0.00E+00&2.82E+07&0.00E+00&17.78&2.14E+10&2.91E+08&1.23E+08&2.62E+10&0.00E+00\\
jython&1.06E+08&0.00E+00&0.00E+00&1.72E+07&0.00E+00&11.41&8.86E+08&1.38E+08&2.80E+07&4.14E+09&0.00E+00\\
luindex&2.77E+05&1.00E+00&1.25E+03&1.00E+01&0.00E+00&5.57&3.74E+07&1.85E+05&8.48E+04&7.81E+07&0.00E+00\\
lusearch-fix&6.32E+06&1.38E+02&9.05E+02&5.12E+02&0.00E+00&85.00&6.60E+08&1.04E+07&4.64E+06&6.28E+08&0.00E+00\\
pmd&3.05E+06&0.00E+00&3.34E+03&4.62E+03&3.00E+00&22.26&4.06E+08&1.04E+07&2.86E+06&1.73E+08&0.00E+00\\
sunflow&1.53E+03&5.00E+00&0.00E+00&0.00E+00&0.00E+00&79.55&1.13E+09&1.71E+08&4.34E+06&4.19E+09&0.00E+00\\
tomcat&2.28E+08&6.04E+02&2.18E+05&7.84E+06&1.93E+05&27.51&1.61E+18&1.07E+08&7.61E+07&4.44E+09&0.00E+00\\
tradesoap&7.31E+08&2.12E+02&1.29E+06&2.39E+06&1.30E+05&64.92&2.96E+10&6.64E+08&2.44E+08&1.50E+10&1.40E+02\\
xalan&2.12E+08&3.48E+02&1.01E+05&0.00E+00&0.00E+00&97.89&5.11E+09&6.12E+07&4.00E+07&3.84E+09&0.00E+00\\
\hline
\multicolumn{12}{c}{\bf ScalaBench}\\
\hdashline[0.5pt/1pt]
apparat&1.35E+07&5.64E+03&5.16E+05&1.19E+06&4.54E+04&15.80&2.69E+10&3.22E+08&2.55E+07&1.00E+11&0.00E+00\\
factorie&3.10E+07&3.00E+00&0.00E+00&9.81E+07&0.00E+00&12.04&1.43E+10&7.43E+09&1.16E+08&6.00E+10&0.00E+00\\
kiama&6.47E+04&0.00E+00&0.00E+00&0.00E+00&0.00E+00&3.12&6.28E+07&9.67E+06&2.10E+06&9.10E+07&0.00E+00\\
scalac&2.52E+06&0.00E+00&0.00E+00&0.00E+00&0.00E+00&15.45&6.36E+08&4.69E+07&6.45E+06&1.27E+09&0.00E+00\\
scaladoc&1.90E+06&0.00E+00&0.00E+00&0.00E+00&0.00E+00&6.11&3.74E+08&3.92E+07&7.62E+06&9.76E+08&0.00E+00\\
scalap&7.83E+04&0.00E+00&0.00E+00&0.00E+00&0.00E+00&2.32&2.05E+07&3.39E+06&3.40E+05&7.73E+07&0.00E+00\\
scalariform&1.90E+06&0.00E+00&0.00E+00&0.00E+00&0.00E+00&15.65&3.12E+08&5.70E+07&4.17E+06&5.78E+08&0.00E+00\\
scalatest&7.83E+05&6.45E+02&1.93E+04&6.53E+04&3.30E+01&20.00&2.56E+08&2.61E+06&7.83E+05&3.51E+07&0.00E+00\\
scalaxb&1.76E+05&0.00E+00&0.00E+00&0.00E+00&0.00E+00&12.42&9.55E+09&1.22E+08&4.08E+06&1.14E+10&0.00E+00\\
specs&1.14E+06&4.10E+01&1.38E+04&9.48E+04&5.10E+01&10.53&3.08E+08&1.33E+07&1.93E+06&1.12E+08&0.00E+00\\
tmt&1.35E+08&5.56E+03&8.70E+01&5.13E+04&5.00E+03&36.54&1.26E+19&3.47E+09&1.75E+07&7.19E+10&0.00E+00\\
\hline
\multicolumn{12}{c}{\bf SPECjvm2008}\\
\hdashline[0.5pt/1pt]
compiler.compiler&4.50E+06&5.00E+00&1.00E+00&2.60E+01&0.00E+00&98.30&1.31E+10&4.17E+08&4.78E+07&1.01E+10&0.00E+00\\
compiler.sunflow&3.31E+07&1.52E+02&1.00E+00&3.00E+01&0.00E+00&97.85&2.38E+10&1.02E+09&1.72E+08&2.98E+10&0.00E+00\\
compress&6.18E+05&4.00E+00&1.00E+00&5.50E+01&0.00E+00&98.56&7.06E+10&2.15E+05&1.43E+05&1.56E+11&0.00E+00\\
crypto.aes&2.94E+07&5.00E+00&1.00E+00&7.70E+01&0.00E+00&97.63&8.29E+09&2.80E+05&3.83E+05&2.91E+09&0.00E+00\\
crypto.rsa&4.10E+07&4.00E+00&1.00E+00&9.74E+02&0.00E+00&97.33&4.17E+09&1.47E+08&1.83E+08&1.92E+09&1.00E+00\\
crypto.signverify&2.68E+09&4.00E+00&1.00E+00&7.30E+01&0.00E+00&97.65&1.71E+10&1.51E+07&2.48E+07&2.55E+10&0.00E+00\\
derby&4.39E+08&1.50E+04&2.60E+07&1.97E+06&1.50E+01&97.92&2.05E+10&1.59E+09&4.25E+08&1.43E+10&0.00E+00\\
mpegaudio&9.38E+06&8.50E+01&3.00E+00&3.19E+03&0.00E+00&98.29&2.86E+10&1.50E+05&4.81E+06&1.73E+10&1.00E+00\\
scimark.fft.large&3.36E+08&6.00E+00&1.00E+00&3.70E+01&0.00E+00&95.31&5.94E+10&4.01E+03&2.84E+03&3.36E+08&0.00E+00\\
scimark.fft.small&3.36E+09&4.00E+00&1.00E+00&3.60E+01&0.00E+00&98.14&3.27E+11&4.90E+05&5.29E+05&3.36E+09&0.00E+00\\
scimark.lu.large&1.34E+08&4.00E+00&1.00E+00&3.20E+01&0.00E+00&92.28&1.01E+11&1.80E+03&1.39E+03&1.34E+08&0.00E+00\\
scimark.lu.small&4.02E+09&4.00E+00&1.00E+00&3.30E+01&0.00E+00&97.83&2.95E+11&6.12E+05&9.16E+05&4.02E+09&0.00E+00\\
scimark.sor.large&5.03E+08&5.00E+00&1.00E+00&3.90E+01&0.00E+00&92.38&1.78E+11&5.34E+03&3.35E+03&5.03E+08&0.00E+00\\
scimark.sor.small&6.00E+08&5.00E+00&1.00E+00&3.90E+01&0.00E+00&98.22&7.24E+10&7.00E+04&5.13E+04&6.01E+08&0.00E+00\\
scimark.sparse.large&3.36E+08&4.00E+00&1.00E+00&3.20E+01&0.00E+00&87.15&1.96E+11&3.64E+03&2.71E+03&3.36E+08&0.00E+00\\
scimark.sparse.small&2.40E+08&5.00E+00&1.00E+00&3.20E+01&0.00E+00&96.42&5.16E+10&2.36E+04&3.33E+04&2.40E+08&0.00E+00\\
serial&2.25E+09&2.20E+02&7.50E+01&8.35E+02&0.00E+00&98.09&3.89E+10&1.78E+09&1.05E+09&3.70E+10&1.00E+00\\
sunflow&1.03E+05&4.49E+02&1.00E+00&5.14E+02&0.00E+00&96.95&1.34E+10&2.54E+09&6.26E+07&6.23E+10&0.00E+00\\
xml.transform&4.76E+08&7.00E+00&1.00E+00&2.40E+01&0.00E+00&97.82&5.80E+09&1.75E+08&7.74E+07&7.73E+09&0.00E+00\\
xml.validation&8.99E+08&5.00E+00&1.00E+00&1.22E+02&0.00E+00&98.80&2.05E+10&5.49E+08&2.11E+08&2.41E+10&0.00E+00\\
\hline
\end{tabular}
\caption{\forcameraready{Unnormalized metrics collected on all analyzed benchmarks.}}
\label{tab:pcametricsext}
\end{table*}

\section{Principal Component Analysis}
\label{sec:app:pca:larger}


\begin{figure*}[t]
\subfigure[PC1 vs PC2.]{
\label{subfig:pca:1vs2:bigger}
\includegraphics[width=\columnwidth]{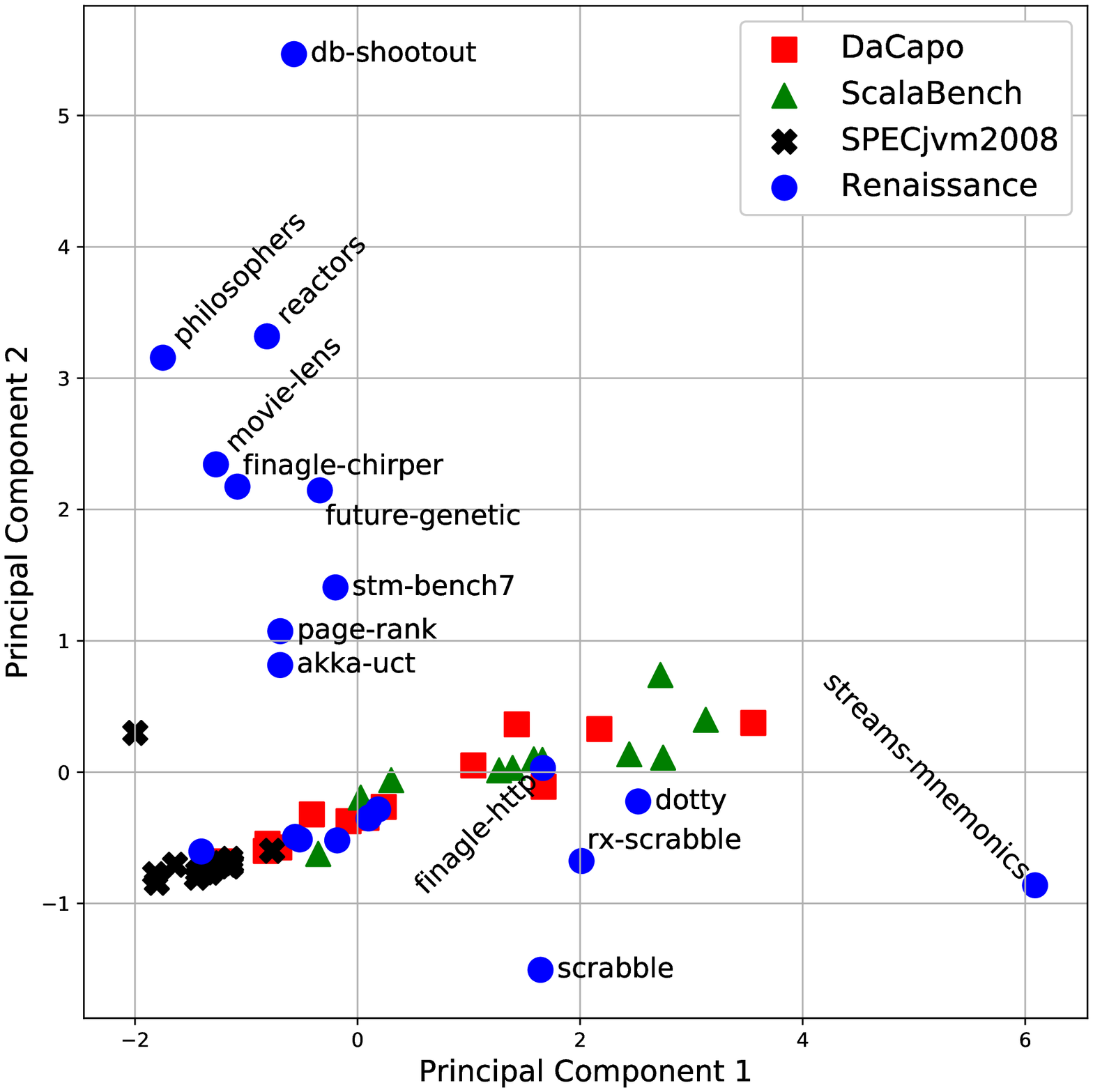}
}
\subfigure[PC3 vs PC4.]{
\label{subfig:pca:3vs4:bigger}
\includegraphics[width=\columnwidth]{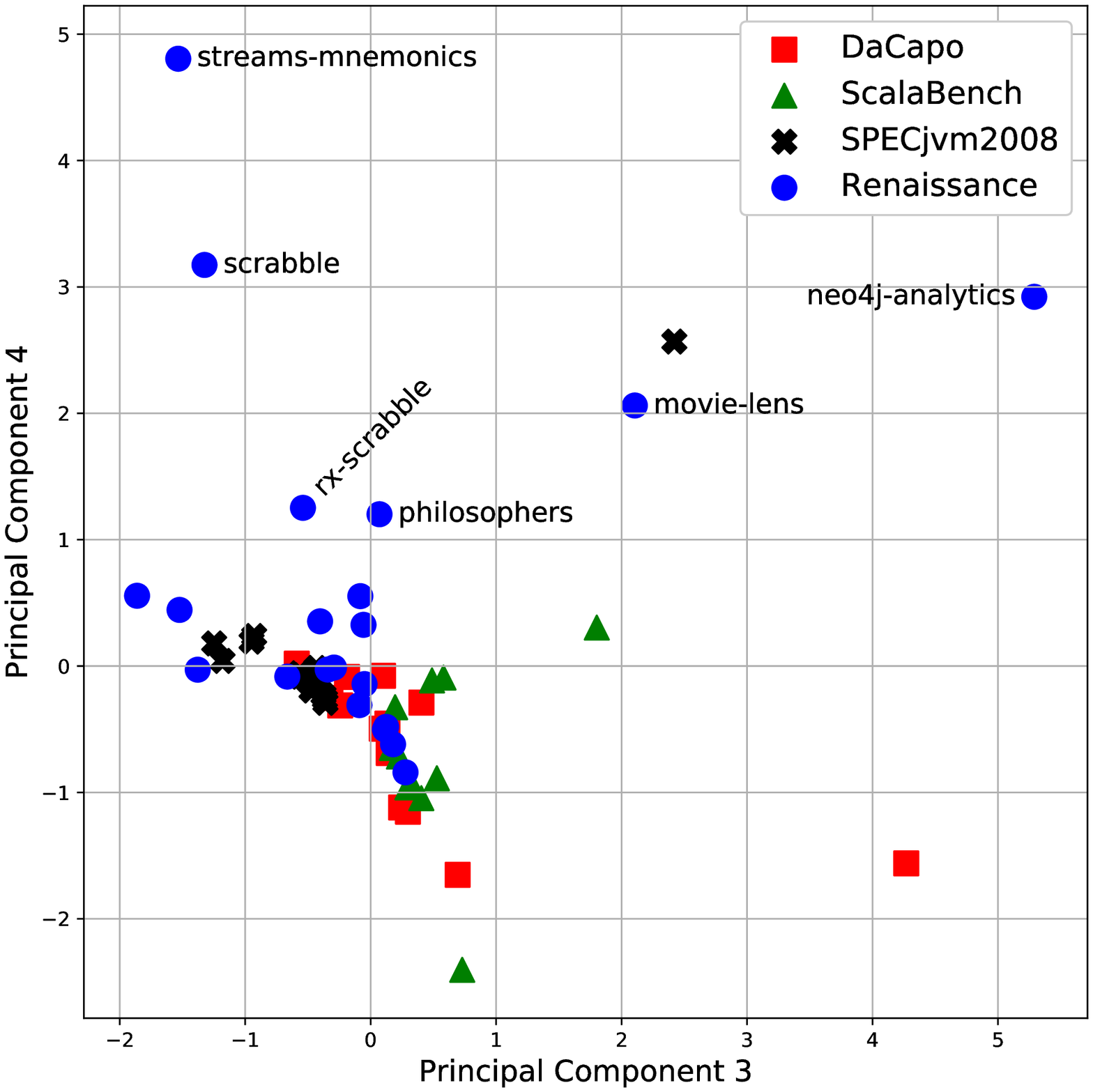}
}
\caption{Scatter plots of benchmark scores over the first four principal components (PCs) - Larger version.}
\label{fig:pca:scatter:larger}
\end{figure*}

In Figure~\ref{fig:pca:scatter:larger}, we report a larger version
of the scatter plots shown in Figure~\ref{fig:pca:scatter}
and discussed in Section~\ref{sec:pca} of the main paper.

%

\section{Additional Data for the Software Complexity Metrics}
\label{sec:app:software-complexity}


In \Cref{tab:chidamber1,tab:chidamber2,tab:chidamber3,tab:chidamber4},
we present additional data for the Chidamber \& Kemerer metrics
(Section~\ref{sec:software-complexity-metrics} in the main paper),
across all four suites.
\Cref{tab:chidamber1,tab:chidamber2} contain the sum for each metric across all benchmarks of a 
suite, while in \Cref{tab:chidamber3,tab:chidamber4} we present the arithmetic mean for each 
metric across all benchmarks of a suite.

\begin{table*}[ht]
	\footnotesize
	\begin{tabular}{lllllll}

		\multicolumn{1}{c}{\bf Benchmark} & 
\multicolumn{1}{r}{\bf WMC} & 
\multicolumn{1}{r}{\bf DIT} & 
\multicolumn{1}{r}{\bf CBO} &
\multicolumn{1}{r}{\bf NOC} & 
\multicolumn{1}{r}{\bf RFC} & 
\multicolumn{1}{r}{\bf LCOM} 
		\\ \toprule

		\multicolumn{7}{c}{\bf Renaissance} \\ \hdashline[0.5pt/1pt]
		
		\multicolumn{1}{l|}{akka-uct}& 
		\multicolumn{1}{r}{34607}& 
		\multicolumn{1}{r}{4938}& 
		\multicolumn{1}{r}{35384}& 
		\multicolumn{1}{r}{2718}& 
		\multicolumn{1}{r}{66665}& 
		\multicolumn{1}{r}{674747} 
		\\
		\multicolumn{1}{l|}{als}& 
		\multicolumn{1}{r}{96524}& 
		\multicolumn{1}{r}{13465}& 
		\multicolumn{1}{r}{95480}& 
		\multicolumn{1}{r}{7488}& 
		\multicolumn{1}{r}{184730}& 
		\multicolumn{1}{r}{5049619} 
		\\
\multicolumn{1}{l|}{chi-square}& 
		\multicolumn{1}{r}{116483}& 
		\multicolumn{1}{r}{13963}& 
		\multicolumn{1}{r}{108544}& 
		\multicolumn{1}{r}{8110}& 
		\multicolumn{1}{r}{232191}& 
		\multicolumn{1}{r}{3465588} 
		\\

		\multicolumn{1}{l|}{db-shootout}& 
		\multicolumn{1}{r}{57652}& 
		\multicolumn{1}{r}{7393}& 
		\multicolumn{1}{r}{51499}& 
		\multicolumn{1}{r}{4285}& 
		\multicolumn{1}{r}{99878}& 
		\multicolumn{1}{r}{1874929} 
		\\
		\multicolumn{1}{l|}{dec-tree}& 
		\multicolumn{1}{r}{206933}& 
		\multicolumn{1}{r}{23936}& 
		\multicolumn{1}{r}{184901}& 
		\multicolumn{1}{r}{14029}& 
		\multicolumn{1}{r}{369131}& 
		\multicolumn{1}{r}{7360650} 
		\\
		\multicolumn{1}{l|}{dotty}& 
		\multicolumn{1}{r}{65887}& 
		\multicolumn{1}{r}{7185}& 
		\multicolumn{1}{r}{62533}& 
		\multicolumn{1}{r}{4121}& 
		\multicolumn{1}{r}{120656}& 
		\multicolumn{1}{r}{1824595} 
		\\
	\multicolumn{1}{l|}{finagle-chirper}& 
		\multicolumn{1}{r}{71465}& 
		\multicolumn{1}{r}{13894}& 
		\multicolumn{1}{r}{78437}& 
		\multicolumn{1}{r}{6322}& 
		\multicolumn{1}{r}{137705}& 
		\multicolumn{1}{r}{1429201} 
		\\

		\multicolumn{1}{l|}{finagle-http}& 
		\multicolumn{1}{r}{65465}& 
		\multicolumn{1}{r}{13122}& 
		\multicolumn{1}{r}{72146}& 
		\multicolumn{1}{r}{5835}& 
		\multicolumn{1}{r}{126281}& 
		\multicolumn{1}{r}{1169093} 
		\\
		\multicolumn{1}{l|}{fj-kmeans}& 
		\multicolumn{1}{r}{22425}& 
		\multicolumn{1}{r}{3092}& 
		\multicolumn{1}{r}{22061}& 
		\multicolumn{1}{r}{1592}& 
		\multicolumn{1}{r}{42584}& 
		\multicolumn{1}{r}{461842} 
		\\
			\multicolumn{1}{l|}{future-genetic}& 
		\multicolumn{1}{r}{26198}& 
		\multicolumn{1}{r}{3499}& 
		\multicolumn{1}{r}{25430}& 
		\multicolumn{1}{r}{1883}& 
		\multicolumn{1}{r}{49263}& 
		\multicolumn{1}{r}{508615} 
		\\
		\multicolumn{1}{l|}{log-regression}& 
		\multicolumn{1}{r}{163424}& 
		\multicolumn{1}{r}{21841}& 
		\multicolumn{1}{r}{161667}& 
		\multicolumn{1}{r}{12057}& 
		\multicolumn{1}{r}{307276}& 
		\multicolumn{1}{r}{5569868} 
		\\

		\multicolumn{1}{l|}{movie-lens}& 
		\multicolumn{1}{r}{101483}& 
		\multicolumn{1}{r}{14335}& 
		\multicolumn{1}{r}{100517}& 
		\multicolumn{1}{r}{8050}& 
		\multicolumn{1}{r}{192950}& 
		\multicolumn{1}{r}{5118756} 
		\\
		\multicolumn{1}{l|}{naive-bayes}& 
		\multicolumn{1}{r}{88885}& 
		\multicolumn{1}{r}{12871}& 
		\multicolumn{1}{r}{91563}& 
		\multicolumn{1}{r}{7130}& 
		\multicolumn{1}{r}{174908}& 
		\multicolumn{1}{r}{1850846} 
		\\
		\multicolumn{1}{l|}{neo4j-analytics}& 
		\multicolumn{1}{r}{119743}& 
		\multicolumn{1}{r}{22172}& 
		\multicolumn{1}{r}{141185}& 
		\multicolumn{1}{r}{11666}& 
		\multicolumn{1}{r}{224669}& 
		\multicolumn{1}{r}{1524820} 
		\\
		\multicolumn{1}{l|}{page-rank}& 
		\multicolumn{1}{r}{93537}& 
		\multicolumn{1}{r}{13939}& 
		\multicolumn{1}{r}{97078}& 
		\multicolumn{1}{r}{7732}& 
		\multicolumn{1}{r}{183346}& 
		\multicolumn{1}{r}{1541349} 
		\\
		\multicolumn{1}{l|}{philosophers}& 
		\multicolumn{1}{r}{24617}& 
		\multicolumn{1}{r}{3432}& 
		\multicolumn{1}{r}{24161}& 
		\multicolumn{1}{r}{1821}& 
		\multicolumn{1}{r}{46714}& 
		\multicolumn{1}{r}{494658} 
		\\
		\multicolumn{1}{l|}{reactors}& 
		\multicolumn{1}{r}{32644}& 
		\multicolumn{1}{r}{4097}& 
		\multicolumn{1}{r}{29610}& 
		\multicolumn{1}{r}{2251}& 
		\multicolumn{1}{r}{60899}& 
		\multicolumn{1}{r}{1066392} 
		\\
		\multicolumn{1}{l|}{rx-scrabble}& 
		\multicolumn{1}{r}{25981}& 
		\multicolumn{1}{r}{3752}& 
		\multicolumn{1}{r}{25829}& 
		\multicolumn{1}{r}{1958}& 
		\multicolumn{1}{r}{49387}& 
		\multicolumn{1}{r}{576353} 
		\\
		\multicolumn{1}{l|}{scrabble}& 
		\multicolumn{1}{r}{24333}& 
		\multicolumn{1}{r}{3380}& 
		\multicolumn{1}{r}{24176}& 
		\multicolumn{1}{r}{1759}& 
		\multicolumn{1}{r}{46212}& 
		\multicolumn{1}{r}{484610} 
		\\

		\multicolumn{1}{l|}{stm-bench7}& 
		\multicolumn{1}{r}{28074}& 
		\multicolumn{1}{r}{3829}& 
		\multicolumn{1}{r}{27159}& 
		\multicolumn{1}{r}{2083}& 
		\multicolumn{1}{r}{52889}& 
		\multicolumn{1}{r}{635890} 
		\\
				\multicolumn{1}{l|}{streams-mnemonics}& 
		\multicolumn{1}{r}{21830}& 
		\multicolumn{1}{r}{3066}& 
		\multicolumn{1}{r}{21757}& 
		\multicolumn{1}{r}{1571}& 
		\multicolumn{1}{r}{41799}& 
		\multicolumn{1}{r}{455958} 
		\\

		\hdashline[0.5pt/1pt]
		
		\multicolumn{1}{l|}{min}& 
		\multicolumn{1}{r}{21830}& 
		\multicolumn{1}{r}{3066}& 
		\multicolumn{1}{r}{21757}& 
		\multicolumn{1}{r}{1571}& 
		\multicolumn{1}{r}{41799}& 
		\multicolumn{1}{r}{455958} 
		\\
		\multicolumn{1}{l|}{max}& 
		\multicolumn{1}{r}{206933}& 
		\multicolumn{1}{r}{23936}& 
		\multicolumn{1}{r}{184901}& 
		\multicolumn{1}{r}{14029}& 
		\multicolumn{1}{r}{369131}& 
		\multicolumn{1}{r}{7360650} 
		\\

		\hdashline[0.5pt/1pt]
		
		\multicolumn{1}{l|}{geomean}& 
		\multicolumn{1}{r}{55533.33}& 
		\multicolumn{1}{r}{7842.22}& 
		\multicolumn{1}{r}{55146.79}& 
		\multicolumn{1}{r}{4212.47}& 
		\multicolumn{1}{r}{105104.97}& 
		\multicolumn{1}{r}{1358042.08} 
		\\

		\bottomrule
	\end{tabular}
	\caption{CK metrics for Renaissance: Sum across all loaded classes of a benchmark.}
	\label{tab:chidamber2}
\end{table*}

\begin{table*}[ht]
	\footnotesize
\begin{tabular}{lllllll}

\multicolumn{1}{c}{\bf Benchmark} & 
\multicolumn{1}{r}{\bf WMC} & 
\multicolumn{1}{r}{\bf DIT} & 
\multicolumn{1}{r}{\bf CBO} &
\multicolumn{1}{r}{\bf NOC} & 
\multicolumn{1}{r}{\bf RFC} & 
\multicolumn{1}{r}{\bf LCOM} 
\\ \toprule

\multicolumn{7}{c}{\bf DaCapo}
\\ \hdashline[0.5pt/1pt]

\multicolumn{1}{l|}{avrora}& 
\multicolumn{1}{r}{13488}& 
\multicolumn{1}{r}{2328}& 
\multicolumn{1}{r}{13719}& 
\multicolumn{1}{r}{1145}& 
\multicolumn{1}{r}{25357}& 
\multicolumn{1}{r}{169973} 
\\
\multicolumn{1}{l|}{batik}& 
\multicolumn{1}{r}{31205}& 
\multicolumn{1}{r}{4958}& 
\multicolumn{1}{r}{30436}& 
\multicolumn{1}{r}{2655}& 
\multicolumn{1}{r}{60275}& 
\multicolumn{1}{r}{333328} 
\\
\multicolumn{1}{l|}{eclipse}& 
\multicolumn{1}{r}{66318}& 
\multicolumn{1}{r}{5753}& 
\multicolumn{1}{r}{45597}& 
\multicolumn{1}{r}{4346}& 
\multicolumn{1}{r}{97757}& 
\multicolumn{1}{r}{600483} 
\\
\multicolumn{1}{l|}{fop}& 
\multicolumn{1}{r}{28162}& 
\multicolumn{1}{r}{4272}& 
\multicolumn{1}{r}{28427}& 
\multicolumn{1}{r}{2278}& 
\multicolumn{1}{r}{56725}& 
\multicolumn{1}{r}{324197} 
\\
\multicolumn{1}{l|}{h2}& 
\multicolumn{1}{r}{20230}& 
\multicolumn{1}{r}{2358}& 
\multicolumn{1}{r}{16492}& 
\multicolumn{1}{r}{1200}& 
\multicolumn{1}{r}{37467}& 
\multicolumn{1}{r}{297210} 
\\
\multicolumn{1}{l|}{jython}& 
\multicolumn{1}{r}{66079}& 
\multicolumn{1}{r}{3595}& 
\multicolumn{1}{r}{34583}& 
\multicolumn{1}{r}{2978}& 
\multicolumn{1}{r}{71881}& 
\multicolumn{1}{r}{616258} 
\\
\multicolumn{1}{l|}{luindex}& 
\multicolumn{1}{r}{14361}& 
\multicolumn{1}{r}{1927}& 
\multicolumn{1}{r}{12737}& 
\multicolumn{1}{r}{1004}& 
\multicolumn{1}{r}{26549}& 
\multicolumn{1}{r}{181134} 
\\
\multicolumn{1}{l|}{lusearch-fix}& 
\multicolumn{1}{r}{12466}& 
\multicolumn{1}{r}{1687}& 
\multicolumn{1}{r}{10640}& 
\multicolumn{1}{r}{857}& 
\multicolumn{1}{r}{22378}& 
\multicolumn{1}{r}{161333} 
\\
\multicolumn{1}{l|}{pmd}& 
\multicolumn{1}{r}{24238}& 
\multicolumn{1}{r}{3406}& 
\multicolumn{1}{r}{21589}& 
\multicolumn{1}{r}{1746}& 
\multicolumn{1}{r}{46074}& 
\multicolumn{1}{r}{526595} 
\\
\multicolumn{1}{l|}{sunflow}& 
\multicolumn{1}{r}{18088}& 
\multicolumn{1}{r}{2359}& 
\multicolumn{1}{r}{17493}& 
\multicolumn{1}{r}{1266}& 
\multicolumn{1}{r}{37288}& 
\multicolumn{1}{r}{245761} 
\\
\multicolumn{1}{l|}{tomcat}& 
\multicolumn{1}{r}{63591}& 
\multicolumn{1}{r}{5347}& 
\multicolumn{1}{r}{37448}& 
\multicolumn{1}{r}{4239}& 
\multicolumn{1}{r}{85921}& 
\multicolumn{1}{r}{616524} 
\\
\multicolumn{1}{l|}{tradebeans}& 
\multicolumn{1}{r}{122123}& 
\multicolumn{1}{r}{5108}& 
\multicolumn{1}{r}{36856}& 
\multicolumn{1}{r}{6347}& 
\multicolumn{1}{r}{74677}& 
\multicolumn{1}{r}{463044} 
\\
\multicolumn{1}{l|}{tradesoap}& 
\multicolumn{1}{r}{124191}& 
\multicolumn{1}{r}{5120}& 
\multicolumn{1}{r}{36971}& 
\multicolumn{1}{r}{6421}& 
\multicolumn{1}{r}{75043}& 
\multicolumn{1}{r}{466569} 
\\

\multicolumn{1}{l|}{xalan}& 
\multicolumn{1}{r}{18203}& 
\multicolumn{1}{r}{2852}& 
\multicolumn{1}{r}{16987}& 
\multicolumn{1}{r}{1364}& 
\multicolumn{1}{r}{35126}& 
\multicolumn{1}{r}{231540} 
\\

\hdashline[0.5pt/1pt]

\multicolumn{1}{l|}{min}& 
\multicolumn{1}{r}{12466}& 
\multicolumn{1}{r}{1687}& 
\multicolumn{1}{r}{10640}& 
\multicolumn{1}{r}{857}& 
\multicolumn{1}{r}{22378}& 
\multicolumn{1}{r}{161333} 
\\
\multicolumn{1}{l|}{max}& 
\multicolumn{1}{r}{124191}& 
\multicolumn{1}{r}{5753}& 
\multicolumn{1}{r}{45597}& 
\multicolumn{1}{r}{6421}& 
\multicolumn{1}{r}{97757}& 
\multicolumn{1}{r}{616524} 
\\

\hdashline[0.5pt/1pt]
\multicolumn{1}{l|}{geomean}& 
\multicolumn{1}{r}{32470}& 
\multicolumn{1}{r}{3376.65}& 
\multicolumn{1}{r}{23275.3}& 
\multicolumn{1}{r}{2160.25}& 
\multicolumn{1}{r}{48461.37}& 
\multicolumn{1}{r}{336191.51} 
\\

\hline

\multicolumn{7}{c}{\bf ScalaBench}
\\ \hdashline[0.5pt/1pt]

\multicolumn{1}{l|}{actors}& 
\multicolumn{1}{r}{41398}& 
\multicolumn{1}{r}{4900}& 
\multicolumn{1}{r}{34981}& 
\multicolumn{1}{r}{2848}& 
\multicolumn{1}{r}{76794}& 
\multicolumn{1}{r}{1216221} 
\\
\multicolumn{1}{l|}{apparat}& 
\multicolumn{1}{r}{34994}& 
\multicolumn{1}{r}{4023}& 
\multicolumn{1}{r}{29838}& 
\multicolumn{1}{r}{2121}& 
\multicolumn{1}{r}{66309}& 
\multicolumn{1}{r}{1344756} 
\\
\multicolumn{1}{l|}{factorie}& 
\multicolumn{1}{r}{24693}& 
\multicolumn{1}{r}{2657}& 
\multicolumn{1}{r}{19713}& 
\multicolumn{1}{r}{1481}& 
\multicolumn{1}{r}{45364}& 
\multicolumn{1}{r}{1080228} 
\\
\multicolumn{1}{l|}{kiama}& 
\multicolumn{1}{r}{31925}& 
\multicolumn{1}{r}{4054}& 
\multicolumn{1}{r}{27664}& 
\multicolumn{1}{r}{2051}& 
\multicolumn{1}{r}{60955}& 
\multicolumn{1}{r}{1259018} 
\\
\multicolumn{1}{l|}{scalac}& 
\multicolumn{1}{r}{57337}& 
\multicolumn{1}{r}{7789}& 
\multicolumn{1}{r}{66036}& 
\multicolumn{1}{r}{4111}& 
\multicolumn{1}{r}{124240}& 
\multicolumn{1}{r}{2298594} 
\\
\multicolumn{1}{l|}{scaladoc}& 
\multicolumn{1}{r}{50655}& 
\multicolumn{1}{r}{6343}& 
\multicolumn{1}{r}{50616}& 
\multicolumn{1}{r}{3379}& 
\multicolumn{1}{r}{103649}& 
\multicolumn{1}{r}{2178195} 
\\
\multicolumn{1}{l|}{scalap}& 
\multicolumn{1}{r}{29661}& 
\multicolumn{1}{r}{3137}& 
\multicolumn{1}{r}{23638}& 
\multicolumn{1}{r}{1743}& 
\multicolumn{1}{r}{54480}& 
\multicolumn{1}{r}{1312560} 
\\
\multicolumn{1}{l|}{scalariform}& 
\multicolumn{1}{r}{32871}& 
\multicolumn{1}{r}{3626}& 
\multicolumn{1}{r}{28467}& 
\multicolumn{1}{r}{2041}& 
\multicolumn{1}{r}{62891}& 
\multicolumn{1}{r}{1352823} 
\\
\multicolumn{1}{l|}{scalatest}& 
\multicolumn{1}{r}{114544}& 
\multicolumn{1}{r}{4589}& 
\multicolumn{1}{r}{37128}& 
\multicolumn{1}{r}{6687}& 
\multicolumn{1}{r}{78209}& 
\multicolumn{1}{r}{1473601} 
\\
\multicolumn{1}{l|}{scalaxb}& 
\multicolumn{1}{r}{30112}& 
\multicolumn{1}{r}{3402}& 
\multicolumn{1}{r}{25528}& 
\multicolumn{1}{r}{1884}& 
\multicolumn{1}{r}{57511}& 
\multicolumn{1}{r}{1180527} 
\\
\multicolumn{1}{l|}{specs}& 
\multicolumn{1}{r}{150895}& 
\multicolumn{1}{r}{6548}& 
\multicolumn{1}{r}{50104}& 
\multicolumn{1}{r}{7890}& 
\multicolumn{1}{r}{105427}& 
\multicolumn{1}{r}{2215152} 
\\
\multicolumn{1}{l|}{tmt}& 
\multicolumn{1}{r}{35875}& 
\multicolumn{1}{r}{3188}& 
\multicolumn{1}{r}{25590}& 
\multicolumn{1}{r}{1941}& 
\multicolumn{1}{r}{64142}& 
\multicolumn{1}{r}{1565153} 
\\

\hdashline[0.5pt/1pt]

\multicolumn{1}{l|}{min}& 
\multicolumn{1}{r}{24693}& 
\multicolumn{1}{r}{2657}& 
\multicolumn{1}{r}{19713}& 
\multicolumn{1}{r}{1481}& 
\multicolumn{1}{r}{45364}& 
\multicolumn{1}{r}{1080228} 
\\
\multicolumn{1}{l|}{max}& 
\multicolumn{1}{r}{150895}& 
\multicolumn{1}{r}{7789}& 
\multicolumn{1}{r}{66036}& 
\multicolumn{1}{r}{7890}& 
\multicolumn{1}{r}{124240}& 
\multicolumn{1}{r}{2298594} 
\\

 \hdashline[0.5pt/1pt]
\multicolumn{1}{l|}{geomean}& 
\multicolumn{1}{r}{44505.06}& 
\multicolumn{1}{r}{4290.27}& 
\multicolumn{1}{r}{32809.23}& 
\multicolumn{1}{r}{2734.69}& 
\multicolumn{1}{r}{71839.71}& 
\multicolumn{1}{r}{1489515.03} 
\\

\hline
\multicolumn{7}{c}{\bf SPECjvm2008}                                                               
\\ \hdashline[0.5pt/1pt]

\multicolumn{1}{l|}{compiler.compiler}& 
\multicolumn{1}{r}{36385}& 
\multicolumn{1}{r}{4744}& 
\multicolumn{1}{r}{36421}& 
\multicolumn{1}{r}{2728}& 
\multicolumn{1}{r}{77428}& 
\multicolumn{1}{r}{598961} 
\\
\multicolumn{1}{l|}{compiler.sunflow}& 
\multicolumn{1}{r}{36456}& 
\multicolumn{1}{r}{4745}& 
\multicolumn{1}{r}{36424}& 
\multicolumn{1}{r}{2728}& 
\multicolumn{1}{r}{77538}& 
\multicolumn{1}{r}{600724} 
\\
\multicolumn{1}{l|}{compress}& 
\multicolumn{1}{r}{30586}& 
\multicolumn{1}{r}{3843}& 
\multicolumn{1}{r}{30009}& 
\multicolumn{1}{r}{2200}& 
\multicolumn{1}{r}{65819}& 
\multicolumn{1}{r}{546396} 
\\
\multicolumn{1}{l|}{crypto.aes}& 
\multicolumn{1}{r}{33789}& 
\multicolumn{1}{r}{4134}& 
\multicolumn{1}{r}{31810}& 
\multicolumn{1}{r}{2413}& 
\multicolumn{1}{r}{69028}& 
\multicolumn{1}{r}{551637} 
\\
\multicolumn{1}{l|}{crypto.rsa}& 
\multicolumn{1}{r}{32724}& 
\multicolumn{1}{r}{4089}& 
\multicolumn{1}{r}{31553}& 
\multicolumn{1}{r}{2381}& 
\multicolumn{1}{r}{68457}& 
\multicolumn{1}{r}{549741} 
\\
\multicolumn{1}{l|}{crypto.signverify}& 
\multicolumn{1}{r}{30999}& 
\multicolumn{1}{r}{3930}& 
\multicolumn{1}{r}{30639}& 
\multicolumn{1}{r}{2250}& 
\multicolumn{1}{r}{66899}& 
\multicolumn{1}{r}{547493} 
\\
\multicolumn{1}{l|}{derby}& 
\multicolumn{1}{r}{55044}& 
\multicolumn{1}{r}{5744}& 
\multicolumn{1}{r}{45745}& 
\multicolumn{1}{r}{3480}& 
\multicolumn{1}{r}{103373}& 
\multicolumn{1}{r}{1131251} 
\\
\multicolumn{1}{l|}{mpegaudio}& 
\multicolumn{1}{r}{31370}& 
\multicolumn{1}{r}{3925}& 
\multicolumn{1}{r}{30483}& 
\multicolumn{1}{r}{2251}& 
\multicolumn{1}{r}{66884}& 
\multicolumn{1}{r}{552206} 
\\
\multicolumn{1}{l|}{scimark.fft.large}& 
\multicolumn{1}{r}{30572}& 
\multicolumn{1}{r}{3835}& 
\multicolumn{1}{r}{29977}& 
\multicolumn{1}{r}{2192}& 
\multicolumn{1}{r}{65767}& 
\multicolumn{1}{r}{546563} 
\\
\multicolumn{1}{l|}{scimark.fft.small}& 
\multicolumn{1}{r}{30572}& 
\multicolumn{1}{r}{3835}& 
\multicolumn{1}{r}{29977}& 
\multicolumn{1}{r}{2192}& 
\multicolumn{1}{r}{65767}& 
\multicolumn{1}{r}{546563} 
\\
\multicolumn{1}{l|}{scimark.lu.large}& 
\multicolumn{1}{r}{30569}& 
\multicolumn{1}{r}{3833}& 
\multicolumn{1}{r}{29968}& 
\multicolumn{1}{r}{2191}& 
\multicolumn{1}{r}{65761}& 
\multicolumn{1}{r}{546558} 
\\
\multicolumn{1}{l|}{scimark.lu.small}& 
\multicolumn{1}{r}{30569}& 
\multicolumn{1}{r}{3832}& 
\multicolumn{1}{r}{29967}& 
\multicolumn{1}{r}{2190}& 
\multicolumn{1}{r}{65761}& 
\multicolumn{1}{r}{546558} 
\\
\multicolumn{1}{l|}{scimark.monte\_carlo}& 
\multicolumn{1}{r}{30560}& 
\multicolumn{1}{r}{3833}& 
\multicolumn{1}{r}{29966}& 
\multicolumn{1}{r}{2191}& 
\multicolumn{1}{r}{65741}& 
\multicolumn{1}{r}{546482} 
\\
\multicolumn{1}{l|}{scimark.sor.large}& 
\multicolumn{1}{r}{30565}& 
\multicolumn{1}{r}{3835}& 
\multicolumn{1}{r}{29975}& 
\multicolumn{1}{r}{2192}& 
\multicolumn{1}{r}{65755}& 
\multicolumn{1}{r}{546493} 
\\
\multicolumn{1}{l|}{scimark.sor.small}& 
\multicolumn{1}{r}{30565}& 
\multicolumn{1}{r}{3835}& 
\multicolumn{1}{r}{29975}& 
\multicolumn{1}{r}{2192}& 
\multicolumn{1}{r}{65755}& 
\multicolumn{1}{r}{546493} 
\\
\multicolumn{1}{l|}{scimark.sparse.large}& 
\multicolumn{1}{r}{30561}& 
\multicolumn{1}{r}{3833}& 
\multicolumn{1}{r}{29967}& 
\multicolumn{1}{r}{2191}& 
\multicolumn{1}{r}{65746}& 
\multicolumn{1}{r}{546486} 
\\
\multicolumn{1}{l|}{scimark.sparse.small}& 
\multicolumn{1}{r}{30561}& 
\multicolumn{1}{r}{3833}& 
\multicolumn{1}{r}{29967}& 
\multicolumn{1}{r}{2191}& 
\multicolumn{1}{r}{65746}& 
\multicolumn{1}{r}{546486} 
\\
\multicolumn{1}{l|}{serial}& 
\multicolumn{1}{r}{32690}& 
\multicolumn{1}{r}{3987}& 
\multicolumn{1}{r}{31200}& 
\multicolumn{1}{r}{2310}& 
\multicolumn{1}{r}{68584}& 
\multicolumn{1}{r}{570658} 
\\
\multicolumn{1}{l|}{sunflow}& 
\multicolumn{1}{r}{31946}& 
\multicolumn{1}{r}{4003}& 
\multicolumn{1}{r}{31463}& 
\multicolumn{1}{r}{2324}& 
\multicolumn{1}{r}{69076}& 
\multicolumn{1}{r}{552092} 
\\
\multicolumn{1}{l|}{xml.transform}& 
\multicolumn{1}{r}{43374}& 
\multicolumn{1}{r}{5654}& 
\multicolumn{1}{r}{40968}& 
\multicolumn{1}{r}{3154}& 
\multicolumn{1}{r}{86981}& 
\multicolumn{1}{r}{625179} 
\\
\multicolumn{1}{l|}{xml.validation}& 
\multicolumn{1}{r}{34578}& 
\multicolumn{1}{r}{4378}& 
\multicolumn{1}{r}{34225}& 
\multicolumn{1}{r}{2545}& 
\multicolumn{1}{r}{74250}& 
\multicolumn{1}{r}{581626} 
\\

\hdashline[0.5pt/1pt]

\multicolumn{1}{l|}{min}& 
\multicolumn{1}{r}{30560}& 
\multicolumn{1}{r}{3832}& 
\multicolumn{1}{r}{29966}& 
\multicolumn{1}{r}{2190}& 
\multicolumn{1}{r}{65741}& 
\multicolumn{1}{r}{546396} 
\\
\multicolumn{1}{l|}{max}& 
\multicolumn{1}{r}{55044}& 
\multicolumn{1}{r}{5744}& 
\multicolumn{1}{r}{45745}& 
\multicolumn{1}{r}{3480}& 
\multicolumn{1}{r}{103373}& 
\multicolumn{1}{r}{1131251} 
\\

\hdashline[0.5pt/1pt]

\multicolumn{1}{l|}{geomean}& 
\multicolumn{1}{r}{33194.67}& 
\multicolumn{1}{r}{4142.17}& 
\multicolumn{1}{r}{32187.12}& 
\multicolumn{1}{r}{2383.19}& 
\multicolumn{1}{r}{70279.58}& 
\multicolumn{1}{r}{578408.18} 
\\

\bottomrule
\end{tabular}
  \caption{
    CK metrics for DaCapo, ScalaBench and SPECjvm2008: Sum across all loaded classes 
    of a benchmark.
  }
  \label{tab:chidamber1}
\end{table*}

\begin{table*}[ht]
	\footnotesize
	\begin{tabular}{lllllll}
		
		\multicolumn{1}{c}{\bf Benchmark} & 
\multicolumn{1}{r}{\bf WMC} & 
\multicolumn{1}{r}{\bf DIT} & 
\multicolumn{1}{r}{\bf CBO} &
\multicolumn{1}{r}{\bf NOC} & 
\multicolumn{1}{r}{\bf RFC} & 
\multicolumn{1}{r}{\bf LCOM} 
		\\ \toprule

		\multicolumn{7}{c}{\bf Renaissance} \\ \hdashline[0.5pt/1pt]
		
		\multicolumn{1}{l|}{akka-uct}& 
		\multicolumn{1}{r}{12.83}& 
		\multicolumn{1}{r}{1.83}& 
		\multicolumn{1}{r}{13.11}& 
		\multicolumn{1}{r}{1.01}& 
		\multicolumn{1}{r}{24.71}& 
		\multicolumn{1}{r}{250.09} 
		\\
		\multicolumn{1}{l|}{als}& 
		\multicolumn{1}{r}{13.51}& 
		\multicolumn{1}{r}{1.88}& 
		\multicolumn{1}{r}{13.36}& 
		\multicolumn{1}{r}{1.05}& 
		\multicolumn{1}{r}{25.85}& 
		\multicolumn{1}{r}{706.54} 
		\\
			\multicolumn{1}{l|}{chi-square}& 
		\multicolumn{1}{r}{14.91}& 
		\multicolumn{1}{r}{1.79}& 
		\multicolumn{1}{r}{13.9}& 
		\multicolumn{1}{r}{1.04}& 
		\multicolumn{1}{r}{29.73}& 
		\multicolumn{1}{r}{443.74} 
		\\

		\multicolumn{1}{l|}{db-shootout}& 
		\multicolumn{1}{r}{14.61}& 
		\multicolumn{1}{r}{1.87}& 
		\multicolumn{1}{r}{13.05}& 
		\multicolumn{1}{r}{1.09}& 
		\multicolumn{1}{r}{25.3}& 
		\multicolumn{1}{r}{475.03} 
		\\
			\multicolumn{1}{l|}{dec-tree}& 
		\multicolumn{1}{r}{16.17}& 
		\multicolumn{1}{r}{1.87}& 
		\multicolumn{1}{r}{14.45}& 
		\multicolumn{1}{r}{1.1}& 
		\multicolumn{1}{r}{28.84}& 
		\multicolumn{1}{r}{575.1} 
		\\
	\multicolumn{1}{l|}{dotty}& 
		\multicolumn{1}{r}{18.48}& 
		\multicolumn{1}{r}{2.02}& 
		\multicolumn{1}{r}{17.54}& 
		\multicolumn{1}{r}{1.16}& 
		\multicolumn{1}{r}{33.84}& 
		\multicolumn{1}{r}{511.81} 
		\\
\multicolumn{1}{l|}{finagle-chirper}& 
		\multicolumn{1}{r}{11.5}& 
		\multicolumn{1}{r}{2.24}& 
		\multicolumn{1}{r}{12.62}& 
		\multicolumn{1}{r}{1.02}& 
		\multicolumn{1}{r}{22.16}& 
		\multicolumn{1}{r}{230} 
		\\
		
		\multicolumn{1}{l|}{finagle-http}& 
		\multicolumn{1}{r}{11.41}& 
		\multicolumn{1}{r}{2.29}& 
		\multicolumn{1}{r}{12.57}& 
		\multicolumn{1}{r}{1.02}& 
		\multicolumn{1}{r}{22}& 
		\multicolumn{1}{r}{203.71} 
		\\
		\multicolumn{1}{l|}{fj-kmeans}& 
		\multicolumn{1}{r}{13.76}& 
		\multicolumn{1}{r}{1.9}& 
		\multicolumn{1}{r}{13.53}& 
		\multicolumn{1}{r}{0.98}& 
		\multicolumn{1}{r}{26.13}& 
		\multicolumn{1}{r}{283.34} 
		\\
		\multicolumn{1}{l|}{future-genetic}& 
		\multicolumn{1}{r}{13.94}& 
		\multicolumn{1}{r}{1.86}& 
		\multicolumn{1}{r}{13.53}& 
		\multicolumn{1}{r}{1}& 
		\multicolumn{1}{r}{26.22}& 
		\multicolumn{1}{r}{270.68} 
		\\
	\multicolumn{1}{l|}{log-regression}& 
		\multicolumn{1}{r}{14.39}& 
		\multicolumn{1}{r}{1.92}& 
		\multicolumn{1}{r}{14.23}& 
		\multicolumn{1}{r}{1.06}& 
		\multicolumn{1}{r}{27.05}& 
		\multicolumn{1}{r}{490.31} 
		\\

		\multicolumn{1}{l|}{movie-lens}& 
		\multicolumn{1}{r}{13.31}& 
		\multicolumn{1}{r}{1.88}& 
		\multicolumn{1}{r}{13.18}& 
		\multicolumn{1}{r}{1.06}& 
		\multicolumn{1}{r}{25.31}& 
		\multicolumn{1}{r}{671.4} 
		\\
		\multicolumn{1}{l|}{naive-bayes}& 
		\multicolumn{1}{r}{13.01}& 
		\multicolumn{1}{r}{1.88}& 
		\multicolumn{1}{r}{13.4}& 
		\multicolumn{1}{r}{1.04}& 
		\multicolumn{1}{r}{25.6}& 
		\multicolumn{1}{r}{270.91} 
		\\
		\multicolumn{1}{l|}{neo4j-analytics}& 
		\multicolumn{1}{r}{11.07}& 
		\multicolumn{1}{r}{2.05}& 
		\multicolumn{1}{r}{13.06}& 
		\multicolumn{1}{r}{1.08}& 
		\multicolumn{1}{r}{20.78}& 
		\multicolumn{1}{r}{141.02} 
		\\
		\multicolumn{1}{l|}{page-rank}& 
		\multicolumn{1}{r}{12.68}& 
		\multicolumn{1}{r}{1.89}& 
		\multicolumn{1}{r}{13.16}& 
		\multicolumn{1}{r}{1.05}& 
		\multicolumn{1}{r}{24.86}& 
		\multicolumn{1}{r}{209} 
		\\
		\multicolumn{1}{l|}{philosophers}& 
		\multicolumn{1}{r}{13.29}& 
		\multicolumn{1}{r}{1.85}& 
		\multicolumn{1}{r}{13.05}& 
		\multicolumn{1}{r}{0.98}& 
		\multicolumn{1}{r}{25.22}& 
		\multicolumn{1}{r}{267.09} 
		\\
		\multicolumn{1}{l|}{reactors}& 
		\multicolumn{1}{r}{14.54}& 
		\multicolumn{1}{r}{1.82}& 
		\multicolumn{1}{r}{13.19}& 
		\multicolumn{1}{r}{1}& 
		\multicolumn{1}{r}{27.13}& 
		\multicolumn{1}{r}{475.01} 
		\\
		\multicolumn{1}{l|}{rx-scrabble}& 
		\multicolumn{1}{r}{13.2}& 
		\multicolumn{1}{r}{1.91}& 
		\multicolumn{1}{r}{13.12}& 
		\multicolumn{1}{r}{0.99}& 
		\multicolumn{1}{r}{25.08}& 
		\multicolumn{1}{r}{292.71} 
		\\
		\multicolumn{1}{l|}{scrabble}& 
		\multicolumn{1}{r}{13.57}& 
		\multicolumn{1}{r}{1.89}& 
		\multicolumn{1}{r}{13.48}& 
		\multicolumn{1}{r}{0.98}& 
		\multicolumn{1}{r}{25.77}& 
		\multicolumn{1}{r}{270.28} 
		\\
		\multicolumn{1}{l|}{stm-bench7}& 
		\multicolumn{1}{r}{13.36}& 
		\multicolumn{1}{r}{1.82}& 
		\multicolumn{1}{r}{12.93}& 
		\multicolumn{1}{r}{0.99}& 
		\multicolumn{1}{r}{25.17}& 
		\multicolumn{1}{r}{302.66} 
		\\
				\multicolumn{1}{l|}{stream-mnemonics}& 
		\multicolumn{1}{r}{13.54}& 
		\multicolumn{1}{r}{1.9}& 
		\multicolumn{1}{r}{13.5}& 
		\multicolumn{1}{r}{0.97}& 
		\multicolumn{1}{r}{25.93}& 
		\multicolumn{1}{r}{282.85} 
		\\

		\hdashline[0.5pt/1pt]
		\multicolumn{1}{l|}{min}& 
		\multicolumn{1}{r}{11.07}& 
		\multicolumn{1}{r}{1.79}& 
		\multicolumn{1}{r}{12.57}& 
		\multicolumn{1}{r}{0.97}& 
		\multicolumn{1}{r}{20.78}& 
		\multicolumn{1}{r}{141.02} 
		\\
		\multicolumn{1}{l|}{max}& 
		\multicolumn{1}{r}{18.48}& 
		\multicolumn{1}{r}{2.29}& 
		\multicolumn{1}{r}{17.54}& 
		\multicolumn{1}{r}{1.16}& 
		\multicolumn{1}{r}{33.84}& 
		\multicolumn{1}{r}{706.54} 
		\\

		\hdashline[0.5pt/1pt]
		
		\multicolumn{1}{l|}{geomean}& 
		\multicolumn{1}{r}{13.58}& 
		\multicolumn{1}{r}{1.92}& 
		\multicolumn{1}{r}{13.49}& 
		\multicolumn{1}{r}{1.03}& 
		\multicolumn{1}{r}{25.71}& 
		\multicolumn{1}{r}{332.19} 
		\\

		\bottomrule
	\end{tabular}
	\caption{CK metrics for Renaissance: Average across all loaded classes of a benchmark.}
	\label{tab:chidamber4}
\end{table*}

\begin{table*}[t]
	\footnotesize
	\begin{tabular}{lllllll}

		\multicolumn{1}{c}{\bf Benchmark} & 
\multicolumn{1}{r}{\bf WMC} & 
\multicolumn{1}{r}{\bf DIT} & 
\multicolumn{1}{r}{\bf CBO} &
\multicolumn{1}{r}{\bf NOC} & 
\multicolumn{1}{r}{\bf RFC} & 
\multicolumn{1}{r}{\bf LCOM} 
		\\ \toprule

		\multicolumn{7}{c}{\bf DaCapo}
		\\ \hdashline[0.5pt/1pt]
		
\multicolumn{1}{l|}{avrora}& 
\multicolumn{1}{r}{11.74}& 
\multicolumn{1}{r}{2.03}& 
\multicolumn{1}{r}{11.94}& 
\multicolumn{1}{r}{1}& 
\multicolumn{1}{r}{22.07}& 
\multicolumn{1}{r}{147.93} 
\\
\multicolumn{1}{l|}{batik}& 
\multicolumn{1}{r}{12.2}& 
\multicolumn{1}{r}{1.94}& 
\multicolumn{1}{r}{11.9}& 
\multicolumn{1}{r}{1.04}& 
\multicolumn{1}{r}{23.57}& 
\multicolumn{1}{r}{130.36} 
\\
\multicolumn{1}{l|}{eclipse}& 
\multicolumn{1}{r}{21.02}& 
\multicolumn{1}{r}{1.82}& 
\multicolumn{1}{r}{14.45}& 
\multicolumn{1}{r}{1.38}& 
\multicolumn{1}{r}{30.98}& 
\multicolumn{1}{r}{190.33} 
\\
\multicolumn{1}{l|}{fop}& 
\multicolumn{1}{r}{12.67}& 
\multicolumn{1}{r}{1.92}& 
\multicolumn{1}{r}{12.79}& 
\multicolumn{1}{r}{1.02}& 
\multicolumn{1}{r}{25.52}& 
\multicolumn{1}{r}{145.84} 
\\
\multicolumn{1}{l|}{h2}& 
\multicolumn{1}{r}{17.42}& 
\multicolumn{1}{r}{2.03}& 
\multicolumn{1}{r}{14.2}& 
\multicolumn{1}{r}{1.03}& 
\multicolumn{1}{r}{32.27}& 
\multicolumn{1}{r}{255.99} 
\\
\multicolumn{1}{l|}{jython}& 
\multicolumn{1}{r}{22.81}& 
\multicolumn{1}{r}{1.24}& 
\multicolumn{1}{r}{11.94}& 
\multicolumn{1}{r}{1.03}& 
\multicolumn{1}{r}{24.81}& 
\multicolumn{1}{r}{212.72} 
\\
\multicolumn{1}{l|}{luindex}& 
\multicolumn{1}{r}{14.68}& 
\multicolumn{1}{r}{1.97}& 
\multicolumn{1}{r}{13.02}& 
\multicolumn{1}{r}{1.03}& 
\multicolumn{1}{r}{27.15}& 
\multicolumn{1}{r}{185.21} 
\\
\multicolumn{1}{l|}{lusearch-fix}& 
\multicolumn{1}{r}{15.13}& 
\multicolumn{1}{r}{2.05}& 
\multicolumn{1}{r}{12.91}& 
\multicolumn{1}{r}{1.04}& 
\multicolumn{1}{r}{27.16}& 
\multicolumn{1}{r}{195.79} 
\\
\multicolumn{1}{l|}{pmd}& 
\multicolumn{1}{r}{14.31}& 
\multicolumn{1}{r}{2.01}& 
\multicolumn{1}{r}{12.74}& 
\multicolumn{1}{r}{1.03}& 
\multicolumn{1}{r}{27.2}& 
\multicolumn{1}{r}{310.86} 
\\
\multicolumn{1}{l|}{sunflow}& 
\multicolumn{1}{r}{13.83}& 
\multicolumn{1}{r}{1.8}& 
\multicolumn{1}{r}{13.37}& 
\multicolumn{1}{r}{0.97}& 
\multicolumn{1}{r}{28.51}& 
\multicolumn{1}{r}{187.89} 
\\
\multicolumn{1}{l|}{tomcat}& 
\multicolumn{1}{r}{22.76}& 
\multicolumn{1}{r}{1.91}& 
\multicolumn{1}{r}{13.4}& 
\multicolumn{1}{r}{1.52}& 
\multicolumn{1}{r}{30.75}& 
\multicolumn{1}{r}{220.66} 
\\
\multicolumn{1}{l|}{tradebeans}& 
\multicolumn{1}{r}{41.89}& 
\multicolumn{1}{r}{1.75}& 
\multicolumn{1}{r}{12.64}& 
\multicolumn{1}{r}{2.18}& 
\multicolumn{1}{r}{25.62}& 
\multicolumn{1}{r}{158.85} 
\\
\multicolumn{1}{l|}{tradesoap}& 
\multicolumn{1}{r}{42.43}& 
\multicolumn{1}{r}{1.75}& 
\multicolumn{1}{r}{12.63}& 
\multicolumn{1}{r}{2.19}& 
\multicolumn{1}{r}{25.64}& 
\multicolumn{1}{r}{159.4} 
\\		
\multicolumn{1}{l|}{xalan}& 
\multicolumn{1}{r}{13.57}& 
\multicolumn{1}{r}{2.13}& 
\multicolumn{1}{r}{12.67}& 
\multicolumn{1}{r}{1.02}& 
\multicolumn{1}{r}{26.19}& 
\multicolumn{1}{r}{172.66} 
\\
	
		\hdashline[0.5pt/1pt]

\multicolumn{1}{l|}{min}& 
\multicolumn{1}{r}{11.74}& 
\multicolumn{1}{r}{1.24}& 
\multicolumn{1}{r}{11.9}& 
\multicolumn{1}{r}{0.97}& 
\multicolumn{1}{r}{22.07}& 
\multicolumn{1}{r}{130.36} 
\\
\multicolumn{1}{l|}{max}& 
\multicolumn{1}{r}{42.43}& 
\multicolumn{1}{r}{2.13}& 
\multicolumn{1}{r}{14.45}& 
\multicolumn{1}{r}{2.19}& 
\multicolumn{1}{r}{32.27}& 
\multicolumn{1}{r}{310.86} 
\\

		\hdashline[0.5pt/1pt]

\multicolumn{1}{l|}{geomean}& 
\multicolumn{1}{r}{17.97}& 
\multicolumn{1}{r}{1.87}& 
\multicolumn{1}{r}{12.88}& 
\multicolumn{1}{r}{1.2}& 
\multicolumn{1}{r}{26.82}& 
\multicolumn{1}{r}{186.05} 
\\

		\hline
		
		\multicolumn{7}{c}{\bf ScalaBench}
		\\ \hdashline[0.5pt/1pt]

\multicolumn{1}{l|}{actors}& 
\multicolumn{1}{r}{15}& 
\multicolumn{1}{r}{1.78}& 
\multicolumn{1}{r}{12.67}& 
\multicolumn{1}{r}{1.03}& 
\multicolumn{1}{r}{27.82}& 
\multicolumn{1}{r}{440.66} 
\\
\multicolumn{1}{l|}{apparat}& 
\multicolumn{1}{r}{16.78}& 
\multicolumn{1}{r}{1.93}& 
\multicolumn{1}{r}{14.31}& 
\multicolumn{1}{r}{1.02}& 
\multicolumn{1}{r}{31.8}& 
\multicolumn{1}{r}{644.97} 
\\
\multicolumn{1}{l|}{factorie}& 
\multicolumn{1}{r}{16.67}& 
\multicolumn{1}{r}{1.79}& 
\multicolumn{1}{r}{13.31}& 
\multicolumn{1}{r}{1}& 
\multicolumn{1}{r}{30.63}& 
\multicolumn{1}{r}{729.39} 
\\
\multicolumn{1}{l|}{kiama}& 
\multicolumn{1}{r}{15.55}& 
\multicolumn{1}{r}{1.97}& 
\multicolumn{1}{r}{13.47}& 
\multicolumn{1}{r}{1}& 
\multicolumn{1}{r}{29.69}& 
\multicolumn{1}{r}{613.26} 
\\
\multicolumn{1}{l|}{scalac}& 
\multicolumn{1}{r}{14.04}& 
\multicolumn{1}{r}{1.91}& 
\multicolumn{1}{r}{16.17}& 
\multicolumn{1}{r}{1.01}& 
\multicolumn{1}{r}{30.43}& 
\multicolumn{1}{r}{562.97} 
\\
\multicolumn{1}{l|}{scaladoc}& 
\multicolumn{1}{r}{15.12}& 
\multicolumn{1}{r}{1.89}& 
\multicolumn{1}{r}{15.1}& 
\multicolumn{1}{r}{1.01}& 
\multicolumn{1}{r}{30.93}& 
\multicolumn{1}{r}{650.01} 
\\
\multicolumn{1}{l|}{scalap}& 
\multicolumn{1}{r}{17.14}& 
\multicolumn{1}{r}{1.81}& 
\multicolumn{1}{r}{13.66}& 
\multicolumn{1}{r}{1.01}& 
\multicolumn{1}{r}{31.47}& 
\multicolumn{1}{r}{758.27} 
\\
\multicolumn{1}{l|}{scalariform}& 
\multicolumn{1}{r}{16.18}& 
\multicolumn{1}{r}{1.79}& 
\multicolumn{1}{r}{14.02}& 
\multicolumn{1}{r}{1}& 
\multicolumn{1}{r}{30.97}& 
\multicolumn{1}{r}{666.09} 
\\
\multicolumn{1}{l|}{scalatest}& 
\multicolumn{1}{r}{42.6}& 
\multicolumn{1}{r}{1.71}& 
\multicolumn{1}{r}{13.81}& 
\multicolumn{1}{r}{2.49}& 
\multicolumn{1}{r}{29.08}& 
\multicolumn{1}{r}{548.01} 
\\
\multicolumn{1}{l|}{scalaxb}& 
\multicolumn{1}{r}{16.04}& 
\multicolumn{1}{r}{1.81}& 
\multicolumn{1}{r}{13.6}& 
\multicolumn{1}{r}{1}& 
\multicolumn{1}{r}{30.64}& 
\multicolumn{1}{r}{628.94} 
\\
\multicolumn{1}{l|}{specs}& 
\multicolumn{1}{r}{40.1}& 
\multicolumn{1}{r}{1.74}& 
\multicolumn{1}{r}{13.31}& 
\multicolumn{1}{r}{2.1}& 
\multicolumn{1}{r}{28.02}& 
\multicolumn{1}{r}{588.67} 
\\
\multicolumn{1}{l|}{tmt}& 
\multicolumn{1}{r}{19.16}& 
\multicolumn{1}{r}{1.7}& 
\multicolumn{1}{r}{13.67}& 
\multicolumn{1}{r}{1.04}& 
\multicolumn{1}{r}{34.26}& 
\multicolumn{1}{r}{836.09} 
\\	
		
		\hdashline[0.5pt/1pt]
		
\multicolumn{1}{l|}{min}& 
\multicolumn{1}{r}{14.04}& 
\multicolumn{1}{r}{1.7}& 
\multicolumn{1}{r}{12.67}& 
\multicolumn{1}{r}{1}& 
\multicolumn{1}{r}{27.82}& 
\multicolumn{1}{r}{440.66} 
\\
\multicolumn{1}{l|}{max}& 
\multicolumn{1}{r}{42.6}& 
\multicolumn{1}{r}{1.97}& 
\multicolumn{1}{r}{16.17}& 
\multicolumn{1}{r}{2.49}& 
\multicolumn{1}{r}{34.26}& 
\multicolumn{1}{r}{836.09} 
\\
		
		\hdashline[0.5pt/1pt]
		
\multicolumn{1}{l|}{geomean}& 
\multicolumn{1}{r}{18.85}& 
\multicolumn{1}{r}{1.82}& 
\multicolumn{1}{r}{13.9}& 
\multicolumn{1}{r}{1.16}& 
\multicolumn{1}{r}{30.43}& 
\multicolumn{1}{r}{631.02} 
\\

		\hline
		\multicolumn{7}{c}{\bf SPECjvm2008}                                                         
		      
		\\ \hdashline[0.5pt/1pt]

\multicolumn{1}{l|}{compiler.compiler}& 
\multicolumn{1}{r}{13.55}& 
\multicolumn{1}{r}{1.77}& 
\multicolumn{1}{r}{13.56}& 
\multicolumn{1}{r}{1.02}& 
\multicolumn{1}{r}{28.83}& 
\multicolumn{1}{r}{222.99} 
\\
\multicolumn{1}{l|}{compiler.sunflow}& 
\multicolumn{1}{r}{13.57}& 
\multicolumn{1}{r}{1.77}& 
\multicolumn{1}{r}{13.56}& 
\multicolumn{1}{r}{1.02}& 
\multicolumn{1}{r}{28.87}& 
\multicolumn{1}{r}{223.65} 
\\
\multicolumn{1}{l|}{compress}& 
\multicolumn{1}{r}{13.83}& 
\multicolumn{1}{r}{1.74}& 
\multicolumn{1}{r}{13.57}& 
\multicolumn{1}{r}{1}& 
\multicolumn{1}{r}{29.77}& 
\multicolumn{1}{r}{247.13} 
\\
\multicolumn{1}{l|}{crypto.aes}& 
\multicolumn{1}{r}{14.32}& 
\multicolumn{1}{r}{1.75}& 
\multicolumn{1}{r}{13.48}& 
\multicolumn{1}{r}{1.02}& 
\multicolumn{1}{r}{29.26}& 
\multicolumn{1}{r}{233.84} 
\\
\multicolumn{1}{l|}{crypto.rsa}& 
\multicolumn{1}{r}{14}& 
\multicolumn{1}{r}{1.75}& 
\multicolumn{1}{r}{13.5}& 
\multicolumn{1}{r}{1.02}& 
\multicolumn{1}{r}{29.29}& 
\multicolumn{1}{r}{235.23} 
\\
\multicolumn{1}{l|}{crypto.signverify}& 
\multicolumn{1}{r}{13.73}& 
\multicolumn{1}{r}{1.74}& 
\multicolumn{1}{r}{13.57}& 
\multicolumn{1}{r}{1}& 
\multicolumn{1}{r}{29.63}& 
\multicolumn{1}{r}{242.47} 
\\
\multicolumn{1}{l|}{derby}& 
\multicolumn{1}{r}{16.9}& 
\multicolumn{1}{r}{1.76}& 
\multicolumn{1}{r}{14.04}& 
\multicolumn{1}{r}{1.07}& 
\multicolumn{1}{r}{31.73}& 
\multicolumn{1}{r}{347.22} 
\\
\multicolumn{1}{l|}{mpegaudio}& 
\multicolumn{1}{r}{13.89}& 
\multicolumn{1}{r}{1.74}& 
\multicolumn{1}{r}{13.49}& 
\multicolumn{1}{r}{1}& 
\multicolumn{1}{r}{29.61}& 
\multicolumn{1}{r}{244.45} 
\\
\multicolumn{1}{l|}{scimark.fft.large}& 
\multicolumn{1}{r}{13.87}& 
\multicolumn{1}{r}{1.74}& 
\multicolumn{1}{r}{13.6}& 
\multicolumn{1}{r}{0.99}& 
\multicolumn{1}{r}{29.84}& 
\multicolumn{1}{r}{247.99} 
\\
\multicolumn{1}{l|}{scimark.fft.small}& 
\multicolumn{1}{r}{13.87}& 
\multicolumn{1}{r}{1.74}& 
\multicolumn{1}{r}{13.6}& 
\multicolumn{1}{r}{0.99}& 
\multicolumn{1}{r}{29.84}& 
\multicolumn{1}{r}{247.99} 
\\
\multicolumn{1}{l|}{scimark.lu.large}& 
\multicolumn{1}{r}{13.88}& 
\multicolumn{1}{r}{1.74}& 
\multicolumn{1}{r}{13.6}& 
\multicolumn{1}{r}{0.99}& 
\multicolumn{1}{r}{29.85}& 
\multicolumn{1}{r}{248.1} 
\\
\multicolumn{1}{l|}{scimark.lu.small}& 
\multicolumn{1}{r}{13.88}& 
\multicolumn{1}{r}{1.74}& 
\multicolumn{1}{r}{13.61}& 
\multicolumn{1}{r}{0.99}& 
\multicolumn{1}{r}{29.86}& 
\multicolumn{1}{r}{248.21} 
\\
\multicolumn{1}{l|}{scimark.monte\_carlo}& 
\multicolumn{1}{r}{13.87}& 
\multicolumn{1}{r}{1.74}& 
\multicolumn{1}{r}{13.6}& 
\multicolumn{1}{r}{0.99}& 
\multicolumn{1}{r}{29.84}& 
\multicolumn{1}{r}{248.06} 
\\
\multicolumn{1}{l|}{scimark.sor.large}& 
\multicolumn{1}{r}{13.87}& 
\multicolumn{1}{r}{1.74}& 
\multicolumn{1}{r}{13.6}& 
\multicolumn{1}{r}{0.99}& 
\multicolumn{1}{r}{29.83}& 
\multicolumn{1}{r}{247.96} 
\\
\multicolumn{1}{l|}{scimark.sor.small}& 
\multicolumn{1}{r}{13.87}& 
\multicolumn{1}{r}{1.74}& 
\multicolumn{1}{r}{13.6}& 
\multicolumn{1}{r}{0.99}& 
\multicolumn{1}{r}{29.83}& 
\multicolumn{1}{r}{247.96} 
\\
\multicolumn{1}{l|}{scimark.sparse.large}& 
\multicolumn{1}{r}{13.87}& 
\multicolumn{1}{r}{1.74}& 
\multicolumn{1}{r}{13.6}& 
\multicolumn{1}{r}{0.99}& 
\multicolumn{1}{r}{29.84}& 
\multicolumn{1}{r}{248.06} 
\\
\multicolumn{1}{l|}{scimark.sparse.small}& 
\multicolumn{1}{r}{13.87}& 
\multicolumn{1}{r}{1.74}& 
\multicolumn{1}{r}{13.6}& 
\multicolumn{1}{r}{0.99}& 
\multicolumn{1}{r}{29.84}& 
\multicolumn{1}{r}{248.06} 
\\
\multicolumn{1}{l|}{serial}& 
\multicolumn{1}{r}{14.15}& 
\multicolumn{1}{r}{1.73}& 
\multicolumn{1}{r}{13.5}& 
\multicolumn{1}{r}{1}& 
\multicolumn{1}{r}{29.68}& 
\multicolumn{1}{r}{246.93} 
\\
\multicolumn{1}{l|}{sunflow}& 
\multicolumn{1}{r}{13.69}& 
\multicolumn{1}{r}{1.72}& 
\multicolumn{1}{r}{13.49}& 
\multicolumn{1}{r}{1}& 
\multicolumn{1}{r}{29.61}& 
\multicolumn{1}{r}{236.64} 
\\
\multicolumn{1}{l|}{xml.transform}& 
\multicolumn{1}{r}{14.31}& 
\multicolumn{1}{r}{1.86}& 
\multicolumn{1}{r}{13.51}& 
\multicolumn{1}{r}{1.04}& 
\multicolumn{1}{r}{28.69}& 
\multicolumn{1}{r}{206.19} 
\\
\multicolumn{1}{l|}{xml.validation}& 
\multicolumn{1}{r}{13.62}& 
\multicolumn{1}{r}{1.72}& 
\multicolumn{1}{r}{13.48}& 
\multicolumn{1}{r}{1}& 
\multicolumn{1}{r}{29.24}& 
\multicolumn{1}{r}{229.08} 
\\

		\hdashline[0.5pt/1pt]

\multicolumn{1}{l|}{min}& 
\multicolumn{1}{r}{13.55}& 
\multicolumn{1}{r}{1.72}& 
\multicolumn{1}{r}{13.48}& 
\multicolumn{1}{r}{0.99}& 
\multicolumn{1}{r}{28.69}& 
\multicolumn{1}{r}{206.19} 
\\
\multicolumn{1}{l|}{max}& 
\multicolumn{1}{r}{16.9}& 
\multicolumn{1}{r}{1.86}& 
\multicolumn{1}{r}{14.04}& 
\multicolumn{1}{r}{1.07}& 
\multicolumn{1}{r}{31.73}& 
\multicolumn{1}{r}{347.22} 
\\

		\hdashline[0.5pt/1pt]

\multicolumn{1}{l|}{geomean}& 
\multicolumn{1}{r}{14}& 
\multicolumn{1}{r}{1.75}& 
\multicolumn{1}{r}{13.58}& 
\multicolumn{1}{r}{1.01}& 
\multicolumn{1}{r}{29.65}& 
\multicolumn{1}{r}{244.03} 
\\

		\bottomrule
	\end{tabular}
	\caption{
		CK metrics for DaCapo, ScalaBench and SPECjvm2008: Average across all loaded classes 
		of a benchmark.
	}
	\label{tab:chidamber3}
\end{table*}

\section{Additional Data for the Optimization Impact Measurements}
\label{sec:app:optimization-impact}


In Tables~\ref{tab:optimization-renaissance} to~\ref{tab:optimization-specjvm},
we provide numerical data for the optimization impact overview from
Figure~\ref{fig:optimization-suite-impact}.
The seven columns -- AC, DS, EAWA, GM, LV, LLC and MHS -- stand for the seven optimizations considered,
namely Atomic-Operation Coalescing, Dominance-Based Duplication Simulation, Escape Analysis with
Atomic Operations, Speculative Guard Motion, Loop Vectorization, Loop-Wide Lock Coarsening, and
Method-Handle Simplification.
In each column, the first number gives the change in benchmark execution times observed when
the relevant optimization is turned off, relative to a baseline with all optimizations
turned on
(positive numbers mean optimization speeds up execution,
negative numbers mean optimization slows down execution).
The second number gives the p-value as computed by the Welch's t-test.

\begin{table*}[ht]
\footnotesize
\centering
\begin{tabular}{r | r l | r l | r l | r l | r l | r l | r l }
\textbf{workload} & \multicolumn{2}{ c |}{\textbf{AC}} & \multicolumn{2}{ c |}{\textbf{DS}} & \multicolumn{2}{ c |}{\textbf{EAWA}} & \multicolumn{2}{ c |}{\textbf{GM}} & \multicolumn{2}{ c |}{\textbf{LV}} & \multicolumn{2}{ c |}{\textbf{LLC}} & \multicolumn{2}{ c |}{\textbf{MHS}} \\ 
  \hline
akka-uct & $+1\%$ & $38\%$ & $+2\%$ & $22\%$ & $+5\%$ & $0\%$ & $+1\%$ & $44\%$ & $+4\%$ & $1\%$ & $+1\%$ & $39\%$ & $+3\%$ & $4\%$ \\ 
  als & $+0\%$ & $81\%$ & $+1\%$ & $33\%$ & $-1\%$ & $14\%$ & $+11\%$ & $0\%$ & $+10\%$ & $0\%$ & $+1\%$ & $29\%$ & $+0\%$ & $82\%$ \\ 
  chi-square & $+4\%$ & $3\%$ & $+4\%$ & $4\%$ & $+5\%$ & $0\%$ & $+5\%$ & $0\%$ & $+3\%$ & $12\%$ & $+2\%$ & $33\%$ & $+4\%$ & $2\%$ \\ 
  db-shootout & $-0\%$ & $35\%$ & $-0\%$ & $63\%$ & $+0\%$ & $14\%$ & $+5\%$ & $0\%$ & $+0\%$ & $24\%$ & $-0\%$ & $29\%$ & $+0\%$ & $48\%$ \\ 
  dec-tree & $+0\%$ & $60\%$ & $+1\%$ & $0\%$ & $-0\%$ & $90\%$ & $+8\%$ & $0\%$ & $+3\%$ & $0\%$ & $-0\%$ & $35\%$ & $-0\%$ & $45\%$ \\ 
  dotty & $+0\%$ & $1\%$ & $+2\%$ & $0\%$ & $+0\%$ & $85\%$ & $+3\%$ & $0\%$ & $+1\%$ & $0\%$ & $+0\%$ & $0\%$ & $+8\%$ & $0\%$ \\ 
  finagle-chirper & $-1\%$ & $90\%$ & $-0\%$ & $96\%$ & $+24\%$ & $0\%$ & $-1\%$ & $88\%$ & $+0\%$ & $91\%$ & $+3\%$ & $23\%$ & $+4\%$ & $18\%$ \\ 
  finagle-http & $-1\%$ & $5\%$ & $+4\%$ & $0\%$ & $-1\%$ & $12\%$ & $+0\%$ & $60\%$ & $-0\%$ & $25\%$ & $-0\%$ & $29\%$ & $-0\%$ & $95\%$ \\ 
  fj-kmeans & $-0\%$ & $16\%$ & $-1\%$ & $0\%$ & $+0\%$ & $90\%$ & $+2\%$ & $0\%$ & $-0\%$ & $6\%$ & $+71\%$ & $0\%$ & $-0\%$ & $62\%$ \\ 
  future-genetic & $+24\%$ & $0\%$ & $+0\%$ & $59\%$ & $+2\%$ & $1\%$ & $+2\%$ & $0\%$ & $+1\%$ & $0\%$ & $+1\%$ & $0\%$ & $+25\%$ & $0\%$ \\ 
  log-regression & $-0\%$ & $89\%$ & $+1\%$ & $58\%$ & $+0\%$ & $73\%$ & $+15\%$ & $0\%$ & $+2\%$ & $6\%$ & $+2\%$ & $1\%$ & $+1\%$ & $28\%$ \\ 
  movie-lens & $+1\%$ & $85\%$ & $+0\%$ & $99\%$ & $+1\%$ & $81\%$ & $+1\%$ & $84\%$ & $-1\%$ & $80\%$ & $-3\%$ & $18\%$ & $+1\%$ & $76\%$ \\ 
  naive-bayes & $+1\%$ & $14\%$ & $-3\%$ & $0\%$ & $+1\%$ & $25\%$ & $+13\%$ & $2\%$ & $+1\%$ & $17\%$ & $+1\%$ & $27\%$ & $-0\%$ & $55\%$ \\ 
  neo4j-analytics & $+0\%$ & $91\%$ & $-4\%$ & $37\%$ & $-7\%$ & $10\%$ & $+5\%$ & $24\%$ & $-3\%$ & $49\%$ & $-0\%$ & $100\%$ & $-4\%$ & $27\%$ \\ 
  page-rank & $-1\%$ & $2\%$ & $-0\%$ & $51\%$ & $-1\%$ & $2\%$ & $+2\%$ & $0\%$ & $-0\%$ & $38\%$ & $-1\%$ & $0\%$ & $-1\%$ & $0\%$ \\ 
  philosophers & $-5\%$ & $5\%$ & $-2\%$ & $32\%$ & $-1\%$ & $43\%$ & $+2\%$ & $9\%$ & $+2\%$ & $22\%$ & $-1\%$ & $64\%$ & $-1\%$ & $62\%$ \\ 
  reactors & $-0\%$ & $42\%$ & $-2\%$ & $0\%$ & $-0\%$ & $11\%$ & $-1\%$ & $3\%$ & $-1\%$ & $1\%$ & $-1\%$ & $4\%$ & $-1\%$ & $16\%$ \\ 
  rx-scrabble & $-0\%$ & $93\%$ & $+1\%$ & $6\%$ & $-0\%$ & $69\%$ & $-1\%$ & $0\%$ & $-1\%$ & $8\%$ & $-0\%$ & $38\%$ & $+1\%$ & $0\%$ \\ 
  scrabble & $+1\%$ & $65\%$ & $+1\%$ & $32\%$ & $-2\%$ & $11\%$ & $+3\%$ & $6\%$ & $-1\%$ & $47\%$ & $-1\%$ & $31\%$ & $+22\%$ & $0\%$ \\ 
  stm-bench7 & $+1\%$ & $21\%$ & $+3\%$ & $0\%$ & $+1\%$ & $26\%$ & $+1\%$ & $6\%$ & $+0\%$ & $12\%$ & $+1\%$ & $1\%$ & $-0\%$ & $96\%$ \\ 
  streams-mnemonics & $+0\%$ & $35\%$ & $+22\%$ & $0\%$ & $+1\%$ & $2\%$ & $+1\%$ & $3\%$ & $+2\%$ & $0\%$ & $+0\%$ & $59\%$ & $+7\%$ & $0\%$ \\ 
   \hline

\end{tabular}
\caption{Optimization impact -- Renaissance benchmarks.}
\label{tab:optimization-renaissance}
\end{table*}

\begin{table*}[ht]
\footnotesize
\centering
\begin{tabular}{r | r l | r l | r l | r l | r l | r l | r l }
\textbf{workload} & \multicolumn{2}{ c |}{\textbf{AC}} & \multicolumn{2}{ c |}{\textbf{DS}} & \multicolumn{2}{ c |}{\textbf{EAWA}} & \multicolumn{2}{ c |}{\textbf{GM}} & \multicolumn{2}{ c |}{\textbf{LV}} & \multicolumn{2}{ c |}{\textbf{LLC}} & \multicolumn{2}{ c |}{\textbf{MHS}} \\ 
  \hline
avrora & $+0\%$ & $3\%$ & $+0\%$ & $4\%$ & $+0\%$ & $90\%$ & $+0\%$ & $17\%$ & $+0\%$ & $19\%$ & $+0\%$ & $0\%$ & $+0\%$ & $7\%$ \\ 
  batik & $-0\%$ & $35\%$ & $-0\%$ & $0\%$ & $+0\%$ & $91\%$ & $+1\%$ & $0\%$ & $+0\%$ & $81\%$ & $-0\%$ & $68\%$ & $-0\%$ & $3\%$ \\ 
  eclipse & $+0\%$ & $26\%$ & $+5\%$ & $0\%$ & $-0\%$ & $39\%$ & $+1\%$ & $0\%$ & $+1\%$ & $0\%$ & $+0\%$ & $10\%$ & $+0\%$ & $11\%$ \\ 
  fop & $+0\%$ & $90\%$ & $+1\%$ & $0\%$ & $+0\%$ & $63\%$ & $+0\%$ & $83\%$ & $+1\%$ & $0\%$ & $-0\%$ & $84\%$ & $+0\%$ & $72\%$ \\ 
  h2 & $+0\%$ & $29\%$ & $+2\%$ & $0\%$ & $-0\%$ & $65\%$ & $+1\%$ & $4\%$ & $+0\%$ & $20\%$ & $+0\%$ & $8\%$ & $+1\%$ & $1\%$ \\ 
  jython & $-1\%$ & $26\%$ & $+5\%$ & $0\%$ & $+1\%$ & $65\%$ & $+2\%$ & $9\%$ & $-0\%$ & $96\%$ & $+1\%$ & $42\%$ & $+0\%$ & $96\%$ \\ 
  luindex & $-0\%$ & $39\%$ & $+3\%$ & $0\%$ & $-0\%$ & $5\%$ & $+2\%$ & $0\%$ & $+0\%$ & $0\%$ & $-0\%$ & $70\%$ & $-1\%$ & $0\%$ \\ 
  lusearch & $-0\%$ & $17\%$ & $+1\%$ & $0\%$ & $-0\%$ & $9\%$ & $-0\%$ & $76\%$ & $-0\%$ & $1\%$ & $-0\%$ & $80\%$ & $-0\%$ & $62\%$ \\ 
  pmd & $-0\%$ & $10\%$ & $-0\%$ & $58\%$ & $+0\%$ & $28\%$ & $-1\%$ & $0\%$ & $-0\%$ & $63\%$ & $+0\%$ & $76\%$ & $+0\%$ & $81\%$ \\ 
  sunflow & $+1\%$ & $1\%$ & $+4\%$ & $0\%$ & $+0\%$ & $78\%$ & $+0\%$ & $24\%$ & $+2\%$ & $2\%$ & $+2\%$ & $1\%$ & $+2\%$ & $1\%$ \\ 
  tomcat & $+0\%$ & $6\%$ & $-0\%$ & $40\%$ & $+0\%$ & $76\%$ & $-0\%$ & $54\%$ & $+0\%$ & $40\%$ & $-0\%$ & $85\%$ & $-0\%$ & $91\%$ \\ 
  tradebeans & $+0\%$ & $19\%$ & $+7\%$ & $0\%$ & $+0\%$ & $46\%$ & $-0\%$ & $33\%$ & $+1\%$ & $0\%$ & $+0\%$ & $78\%$ & $+0\%$ & $85\%$ \\ 
  tradesoap & $+3\%$ & $1\%$ & $-2\%$ & $0\%$ & $-2\%$ & $4\%$ & $-0\%$ & $80\%$ & $+1\%$ & $35\%$ & $+0\%$ & $70\%$ & $-3\%$ & $0\%$ \\ 
  xalan & $+1\%$ & $4\%$ & $+1\%$ & $0\%$ & $+0\%$ & $52\%$ & $+0\%$ & $1\%$ & $+0\%$ & $6\%$ & $+0\%$ & $42\%$ & $+0\%$ & $2\%$ \\ 
   \hline

\end{tabular}
\caption{Optimization impact -- DaCapo benchmarks.}
\label{tab:optimization-dacapo}
\end{table*}

\begin{table*}[ht]
\footnotesize
\centering
\begin{tabular}{r | r l | r l | r l | r l | r l | r l | r l }
\textbf{workload} & \multicolumn{2}{ c |}{\textbf{AC}} & \multicolumn{2}{ c |}{\textbf{DS}} & \multicolumn{2}{ c |}{\textbf{EAWA}} & \multicolumn{2}{ c |}{\textbf{GM}} & \multicolumn{2}{ c |}{\textbf{LV}} & \multicolumn{2}{ c |}{\textbf{LLC}} & \multicolumn{2}{ c |}{\textbf{MHS}} \\ 
  \hline
actors & $+0\%$ & $51\%$ & $+1\%$ & $0\%$ & $+1\%$ & $0\%$ & $+0\%$ & $35\%$ & $+0\%$ & $5\%$ & $-0\%$ & $86\%$ & $+0\%$ & $4\%$ \\ 
  apparat & $+1\%$ & $2\%$ & $-1\%$ & $14\%$ & $-1\%$ & $19\%$ & $+0\%$ & $83\%$ & $+1\%$ & $10\%$ & $-0\%$ & $62\%$ & $-0\%$ & $74\%$ \\ 
  factorie & $+2\%$ & $0\%$ & $+7\%$ & $0\%$ & $+1\%$ & $1\%$ & $-2\%$ & $0\%$ & $+1\%$ & $30\%$ & $+1\%$ & $7\%$ & $+1\%$ & $27\%$ \\ 
  kiama & $-0\%$ & $37\%$ & $+4\%$ & $0\%$ & $-0\%$ & $24\%$ & $+1\%$ & $0\%$ & $+1\%$ & $0\%$ & $+0\%$ & $24\%$ & $+0\%$ & $60\%$ \\ 
  scalac & $-0\%$ & $77\%$ & $+1\%$ & $0\%$ & $+0\%$ & $38\%$ & $-0\%$ & $96\%$ & $+0\%$ & $20\%$ & $-0\%$ & $32\%$ & $-0\%$ & $10\%$ \\ 
  scaladoc & $-2\%$ & $0\%$ & $+0\%$ & $65\%$ & $-3\%$ & $0\%$ & $-2\%$ & $0\%$ & $-1\%$ & $23\%$ & $-1\%$ & $10\%$ & $-1\%$ & $40\%$ \\ 
  scalap & $-0\%$ & $1\%$ & $+1\%$ & $0\%$ & $-0\%$ & $0\%$ & $+9\%$ & $0\%$ & $+2\%$ & $0\%$ & $-0\%$ & $7\%$ & $-0\%$ & $0\%$ \\ 
  scalariform & $+0\%$ & $5\%$ & $+1\%$ & $0\%$ & $-0\%$ & $49\%$ & $+0\%$ & $1\%$ & $+0\%$ & $2\%$ & $+0\%$ & $64\%$ & $-0\%$ & $22\%$ \\ 
  scalatest & $+0\%$ & $90\%$ & $-1\%$ & $19\%$ & $-1\%$ & $34\%$ & $+0\%$ & $83\%$ & $+1\%$ & $2\%$ & $+1\%$ & $41\%$ & $+0\%$ & $100\%$ \\ 
  scalaxb & $+1\%$ & $88\%$ & $+8\%$ & $5\%$ & $+1\%$ & $87\%$ & $+6\%$ & $15\%$ & $+6\%$ & $18\%$ & $+7\%$ & $8\%$ & $+2\%$ & $72\%$ \\ 
  specs & $-0\%$ & $16\%$ & $+0\%$ & $18\%$ & $-0\%$ & $11\%$ & $+0\%$ & $2\%$ & $+0\%$ & $78\%$ & $-0\%$ & $20\%$ & $-0\%$ & $45\%$ \\ 
  tmt & $+0\%$ & $10\%$ & $+1\%$ & $0\%$ & $+0\%$ & $0\%$ & $+13\%$ & $0\%$ & $+1\%$ & $0\%$ & $+0\%$ & $6\%$ & $+0\%$ & $42\%$ \\ 
   \hline

\end{tabular}
\caption{Optimization impact -- ScalaBench benchmarks.}
\label{tab:optimization-scalabench}
\end{table*}

\begin{table*}[ht]
\footnotesize
\centering
\begin{tabular}{r | r l | r l | r l | r l | r l | r l | r l }
\textbf{workload} & \multicolumn{2}{ c |}{\textbf{AC}} & \multicolumn{2}{ c |}{\textbf{DS}} & \multicolumn{2}{ c |}{\textbf{EAWA}} & \multicolumn{2}{ c |}{\textbf{GM}} & \multicolumn{2}{ c |}{\textbf{LV}} & \multicolumn{2}{ c |}{\textbf{LLC}} & \multicolumn{2}{ c |}{\textbf{MHS}} \\ 
  \hline
compiler.compiler & $+0\%$ & $8\%$ & $+1\%$ & $0\%$ & $-0\%$ & $10\%$ & $+3\%$ & $0\%$ & $+1\%$ & $0\%$ & $-0\%$ & $35\%$ & $-0\%$ & $73\%$ \\ 
  compiler.sunflow & $-0\%$ & $42\%$ & $+1\%$ & $0\%$ & $+0\%$ & $39\%$ & $+2\%$ & $0\%$ & $+1\%$ & $0\%$ & $-0\%$ & $39\%$ & $+0\%$ & $12\%$ \\ 
  compress & $-0\%$ & $33\%$ & $-2\%$ & $0\%$ & $+0\%$ & $77\%$ & $+2\%$ & $0\%$ & $+4\%$ & $0\%$ & $-0\%$ & $82\%$ & $-0\%$ & $35\%$ \\ 
  crypto.aes & $-0\%$ & $5\%$ & $-0\%$ & $67\%$ & $-0\%$ & $37\%$ & $+1\%$ & $0\%$ & $+1\%$ & $0\%$ & $-0\%$ & $8\%$ & $-0\%$ & $4\%$ \\ 
  crypto.rsa & $-0\%$ & $20\%$ & $+0\%$ & $2\%$ & $-0\%$ & $36\%$ & $+0\%$ & $77\%$ & $-0\%$ & $34\%$ & $-0\%$ & $29\%$ & $-0\%$ & $15\%$ \\ 
  crypto.signverify & $-0\%$ & $92\%$ & $+0\%$ & $69\%$ & $-0\%$ & $64\%$ & $+9\%$ & $0\%$ & $-0\%$ & $74\%$ & $-0\%$ & $100\%$ & $+0\%$ & $87\%$ \\ 
  derby & $+0\%$ & $44\%$ & $+0\%$ & $59\%$ & $-0\%$ & $48\%$ & $-1\%$ & $1\%$ & $-1\%$ & $18\%$ & $+0\%$ & $58\%$ & $+0\%$ & $72\%$ \\ 
  mpegaudio & $-0\%$ & $84\%$ & $-3\%$ & $0\%$ & $+0\%$ & $26\%$ & $+5\%$ & $0\%$ & $+0\%$ & $69\%$ & $+0\%$ & $50\%$ & $+0\%$ & $31\%$ \\ 
  scimark.fft.large & $-3\%$ & $1\%$ & $-2\%$ & $3\%$ & $-3\%$ & $2\%$ & $-1\%$ & $19\%$ & $-3\%$ & $0\%$ & $-2\%$ & $7\%$ & $-1\%$ & $49\%$ \\ 
  scimark.fft.small & $-1\%$ & $44\%$ & $+2\%$ & $33\%$ & $-3\%$ & $9\%$ & $-1\%$ & $65\%$ & $-2\%$ & $22\%$ & $-3\%$ & $4\%$ & $-1\%$ & $68\%$ \\ 
  scimark.lu.large & $-0\%$ & $11\%$ & $-0\%$ & $57\%$ & $-0\%$ & $8\%$ & $+69\%$ & $0\%$ & $+29\%$ & $0\%$ & $-0\%$ & $6\%$ & $+0\%$ & $81\%$ \\ 
  scimark.lu.small & $+0\%$ & $40\%$ & $+1\%$ & $0\%$ & $+0\%$ & $16\%$ & $+137\%$ & $0\%$ & $+58\%$ & $0\%$ & $+0\%$ & $92\%$ & $+0\%$ & $1\%$ \\ 
  scimark.monte\_carlo & $+2\%$ & $30\%$ & $+7\%$ & $0\%$ & $-0\%$ & $83\%$ & $-0\%$ & $83\%$ & $+0\%$ & $89\%$ & $+1\%$ & $61\%$ & $+1\%$ & $62\%$ \\ 
  scimark.sor.large & $+0\%$ & $4\%$ & $-0\%$ & $21\%$ & $+0\%$ & $0\%$ & $+34\%$ & $0\%$ & $-0\%$ & $25\%$ & $+0\%$ & $13\%$ & $-0\%$ & $44\%$ \\ 
  scimark.sor.small & $-0\%$ & $64\%$ & $-0\%$ & $44\%$ & $+0\%$ & $65\%$ & $+36\%$ & $0\%$ & $+0\%$ & $20\%$ & $-0\%$ & $32\%$ & $+0\%$ & $38\%$ \\ 
  scimark.sparse.large & $+0\%$ & $4\%$ & $+1\%$ & $0\%$ & $+0\%$ & $4\%$ & $+16\%$ & $0\%$ & $+0\%$ & $2\%$ & $+0\%$ & $46\%$ & $+0\%$ & $16\%$ \\ 
  scimark.sparse.small & $-0\%$ & $2\%$ & $-0\%$ & $0\%$ & $-0\%$ & $6\%$ & $-10\%$ & $0\%$ & $-0\%$ & $0\%$ & $+0\%$ & $1\%$ & $-0\%$ & $6\%$ \\ 
  serial & $+0\%$ & $94\%$ & $+2\%$ & $0\%$ & $+1\%$ & $4\%$ & $+4\%$ & $0\%$ & $+1\%$ & $5\%$ & $-1\%$ & $11\%$ & $+0\%$ & $39\%$ \\ 
  sunflow & $+1\%$ & $32\%$ & $+2\%$ & $1\%$ & $+1\%$ & $19\%$ & $+1\%$ & $17\%$ & $+2\%$ & $1\%$ & $+1\%$ & $29\%$ & $+1\%$ & $16\%$ \\ 
  xml.transform & $+0\%$ & $73\%$ & $+2\%$ & $0\%$ & $-0\%$ & $60\%$ & $+3\%$ & $0\%$ & $+0\%$ & $24\%$ & $+0\%$ & $83\%$ & $+0\%$ & $54\%$ \\ 
  xml.validation & $-1\%$ & $0\%$ & $+1\%$ & $6\%$ & $-1\%$ & $10\%$ & $-1\%$ & $13\%$ & $-1\%$ & $1\%$ & $-1\%$ & $2\%$ & $-1\%$ & $5\%$ \\ 
   \hline

\end{tabular}
\caption{Optimization impact -- SPECjvm2008 benchmarks.}
\label{tab:optimization-specjvm}
\end{table*}

\forcameraready{
Table~\ref{tab:compilation} provides estimate on the compilation overhead associated with each of the seven optimizations considered.
In each row, the value gives the relative reduction in compiler thread execution time when the particular optimization is disabled,
measured over the entire warm up period. The values are aggregated across all benchmarks.

\begin{table*}[ht]
\footnotesize
\centering
\begin{tabular}{l r}
\textbf{optimization} & \textbf{compilation time change} \\ 
  \hline
Atomic-Operation Coalescing & 0.6\% \\ 
  Dominance-Based Duplication Simulation & 19.6\% \\ 
  Loop-Wide Lock Coarsening & 6.7\% \\ 
  Method-Handle Simplification & 7.2\% \\ 
  Speculative Guard Motion & 5.8\% \\ 
  Loop Vectorization & 5.1\% \\ 
  Escape Analysis with Atomic Operations & 6.9\% \\ 
   \hline

\end{tabular}
\caption{Compilation time associated with individual optimizations.}
\label{tab:compilation}
\end{table*}
}

\section{Related Work}
\label{sec:app:related}


Since its introduction in 2006,
the \mbox{DaCapo} suite~\cite{Blackburn:2006:DBJ:1167515.1167488}
has been a de facto standard for JVM benchmarking.
While much of the original motivation for the DaCapo suite
was to understand object and memory behavior in complex Java applications,
this suite is still actively used to evaluate not only JVM components
such as JIT compilers~\cite{Prokopec:2019:OII:3314872.3314893,Leopoldseder:2018:DDS:3179541.3168811,Prokopec:2017:MCO:3136000.3136002,Eisl:2016:TRA:2972206.2972211,Stadler:2014:PEA:2581122.2544157}
and garbage collectors~\cite{Akram:2018:WGC:3192366.3192392,Nguyen:2016:YHB:3026877.3026905},
but also tools such as profilers~\cite{Schorgenhumer:2017:ESL:3030207.3030234,Bruno:2017:PAP:3135974.3135986},
data-race detectors~\cite{Biswas:2017:LDR:3033019.3033020,Wood:2017:IBD:3152284.3133893},
memory monitors and contention analyzers~\cite{Weninger:2018:UCM:3184407.3184412,Hofer:2016:ETV:2851553.2851559},
static analyzers~\cite{Grech:2018:SHU:3213846.3213860,Thiessen:2017:CTP:3062341.3062359},
and debuggers~\cite{Liu:2018:DFC:3192366.3192390}.

The subsequently proposed
\mbox{ScalaBench} suite~\cite{Sewe:2011:DCC:2048066.2048118,Sewe:2012}
identified a range of typical Scala programs,
and argued that Scala and Java programs have considerably different
distributions of instructions,
polymorphic calls, object allocations, and method sizes.
This observation that benchmark suites tend to over-represent
certain programming styles was also noticed in other languages,
(e.g., JavaScript~\cite{Ratanaworabhan:2010:JCB:1863166.1863169}).
On the other hand,
the \mbox{SPECjvm2008} benchmark suite~\cite{specjvm2008}
focused more on the core Java functionality.
Most of the \mbox{SPECjvm2008} benchmarks are considerably smaller
than the \mbox{DaCapo} and \mbox{ScalaBench} benchmarks,
and do not use a lot of object-oriented abstractions --
\mbox{SPECjvm2008} exercises classic JIT compiler optimizations,
such as instruction scheduling and loop optimizations~\cite{duboscq-thesis}.

The tuning of compilers such as C2~\cite{Paleczny:2001:JHT:1267847.1267848}
and Graal~\cite{graalvm-download,Duboscq:2013:IRS:2542142.2542143}
was heavily influenced by the \mbox{DaCapo}, \mbox{ScalaBench}, and \mbox{SPECjvm2008} suites.
Given that these existing benchmark suites do not exercise
many frameworks and language extensions that gained popularity in the recent years,
we looked for workloads exercising frameworks such as
Java Streams~\cite{jep107} and Parallel Collections~\cite{prokopec11,7092728,Prokopec:196627},
Reactive Extensions~\cite{rx},
Akka~\cite{akka},
Scala actors~\cite{Haller:2007:AUT:1764606.1764620}
and Reactors~\cite{Prokopec:2015:ICE:2814228.2814245,Prokopec:2016:PSR:3001886.3001891,Prokopec:2017:EBB:3133850.3133865,Prokopec:2014:CAM:2637647.2637656},
coroutines~\cite{kotlin-coroutines,DBLP:conf/ecoop/ProkopecL18,coroutines-tech-report},
Apache~Spark~\cite{Zaharia:2010:SCC:1863103.1863113},
futures and promises~\cite{SIP14},
Netty~\cite{github-netty},
Twitter~Finagle~\cite{github-finagle},
and Neo4J~\cite{github-neo4j}.
Most of these frameworks either assist in structuring concurrent programs,
or enable programmers to declaratively specify data-parallel processing tasks.
In both cases, they achieve these goals by providing a higher level of abstraction --
for example, Finagle supports functional-style composition of future values,
while Apache~Spark exposes data-processing combinators for distributed datasets.
By inspecting the IR of the open-source Graal compiler
(c.f. Section~\ref{sec:compiler-optimizations}),
we found that many of the benchmarks
exercise the interaction between different types of JIT compiler optimizations:
optimizations, such as inlining, duplication~\cite{Leopoldseder:2018:DDS:3179541.3168811},
and partial escape analysis~\cite{Stadler:2014:PEA:2581122.2544157},
typically start by reducing the level of abstraction in these frameworks,
and then trigger more low-level optimizations such as
guard motion~\cite{duboscq-thesis}, vectorization, or atomic-operation coalescing.
Aside from a challenge in dealing with high-level abstractions,
the new concurrency primitives in modern benchmarks pose new optimization opportunities,
such as contention elimination~\cite{Lu:2001:CER:379539.379568},
application-specific work-stealing~\cite{10.1007/978-3-319-09967-54},
NUMA-aware node replication~\cite{Calciu:2017:BCD:3037697.3037721},
speculative spinning~\cite{Prokopec2017},
access path caching~\cite{Prokopec:2018:CCL:3178487.3178498,Prokopec2018,techreport-17-prokopec,cache-trie-remove-dataset},
or other traditional compiler optimizations applied to concurrent programs~\cite{Sevcik:2011:SOS:1993498.1993534}.
Many of these newer optimizations may be applicable to domains
such as concurrent data structures, which have been extensively studied
on the JVM~\cite{Oshman:2013:SLC:2484239.2484270,Prokopec2016,cliffclick,prokopec12ctries,EPFL-REPORT-166908,Prokopec2011,Basin:2017:KKM:3018743.3018761,Prokopec:2015:SLQ:2774975.2774976,bronsonavl,prokopec12flowpools,dougleahome}.

Unlike some other suites whose goal
was to simulate deployment in clusters and Cloud environments,
such as CloudSuite~\cite{Ferdman:2012:CCS:2248487.2150982},
our design decision was to follow the philosophy of \mbox{DaCapo} and \mbox{ScalaBench},
in which benchmarks are executed within a single JVM instance,
whose execution characteristics can be understood more easily.
Still, we found some alternative suites useful:
for example, we took the ¡movie-lens¡ benchmark for Apache~Spark from CloudSuite,
and we adapted it to use Spark's single-process mode.

Several other benchmarks were either inspired by or adapted from existing workloads.
The ¡naive-bayes¡, ¡log-regression¡, ¡als¡, ¡dec-tree¡ and ¡chi-square¡ benchmarks
directly work with several machine-learning algorithms from Apache~Spark MLLib,
and some of these benchmarks were inspired by the \mbox{SparkPerf} suite~\cite{spark-perf}.
The \emph{Shakespeare plays Scrabble} benchmark~\cite{scrabble-benchmark}
was presented by José Paumard at the Virtual Technology Summit~2015
to demonstrate an advanced usage of Java Streams,
and we directly adopted it as our ¡scrabble¡ benchmark.
The ¡rx-scrabble¡ is a version of the ¡scrabble¡ benchmark
that uses the Reactive Extensions framework instead of Java Streams.
The ¡streams-mnemonics¡ benchmark is rewritten
from the \emph{Phone Mnemonics} benchmark that was originally used
to demonstrate the usage of Scala collections~\cite{state-of-scala}.
The ¡stm-bench7¡ benchmark is STMBench7~\cite{Guerraoui:2007:SBS:1272996.1273029}
applied to ScalaSTM~\cite{scala-stm-expert-group,Bronson10ccstm:a},
a software transactional memory implementation for Scala,
while the ¡philosophers¡ benchmark is
ScalaSTM's \emph{Reality-Show Philosophers} usage example.

\balance
\bibliography{main}


\begin{thebibliography}{78}


\ifx \showCODEN    \undefined \def \showCODEN     #1{\unskip}     \fi
\ifx \showDOI      \undefined \def \showDOI       #1{#1}\fi
\ifx \showISBNx    \undefined \def \showISBNx     #1{\unskip}     \fi
\ifx \showISBNxiii \undefined \def \showISBNxiii  #1{\unskip}     \fi
\ifx \showISSN     \undefined \def \showISSN      #1{\unskip}     \fi
\ifx \showLCCN     \undefined \def \showLCCN      #1{\unskip}     \fi
\ifx \shownote     \undefined \def \shownote      #1{#1}          \fi
\ifx \showarticletitle \undefined \def \showarticletitle #1{#1}   \fi
\ifx \showURL      \undefined \def \showURL       {\relax}        \fi
\providecommand\bibfield[2]{#2}
\providecommand\bibinfo[2]{#2}
\providecommand\natexlab[1]{#1}
\providecommand\showeprint[2][]{arXiv:#2}

\bibitem[\protect\citeauthoryear{??}{spe}{2008}]%
        {specjvm2008}
 \bibinfo{year}{2008}\natexlab{}.
\newblock \bibinfo{title}{{SPECjvm2008}}.
\newblock
\newblock
\newblock
\shownote{\url{https://www.spec.org/jvm2008/}.}


\bibitem[\protect\citeauthoryear{??}{akk}{2018}]%
        {akka}
 \bibinfo{year}{2018}\natexlab{}.
\newblock \bibinfo{title}{Akka Documentation}.
\newblock
\newblock
\newblock
\shownote{\url{http://akka.io/docs/}.}


\bibitem[\protect\citeauthoryear{??}{gra}{2018}]%
        {graalvm-download}
 \bibinfo{year}{2018}\natexlab{}.
\newblock \bibinfo{title}{{GraalVM} {Website}}.
\newblock
\newblock
\urldef\tempurl%
\url{https://www.graalvm.org/downloads/}
\showURL{%
\tempurl}


\bibitem[\protect\citeauthoryear{??}{kot}{2018}]%
        {kotlin-coroutines}
 \bibinfo{year}{2018}\natexlab{}.
\newblock \bibinfo{title}{Kotlin {Coroutines}}.
\newblock
  \bibinfo{howpublished}{\url{https://github.com/Kotlin/kotlinx.coroutines/blob/master/coroutines-guide.md}}.
\newblock
\newblock
\shownote{Accessed: 2018-11-15.}


\bibitem[\protect\citeauthoryear{??}{git}{2018a}]%
        {github-jenetics}
 \bibinfo{year}{2018}\natexlab{a}.
\newblock \bibinfo{title}{Open-Source {Jenetics} Repository at {GitHub}}.
\newblock
\newblock
\newblock
\shownote{\url{https://github.com/jenetics/jenetics}.}


\bibitem[\protect\citeauthoryear{??}{git}{2018b}]%
        {github-neo4j}
 \bibinfo{year}{2018}\natexlab{b}.
\newblock \bibinfo{title}{Open-Source {Neo4J} Repository at {GitHub}}.
\newblock
\newblock
\newblock
\shownote{\url{https://github.com/neo4j/neo4j}.}


\bibitem[\protect\citeauthoryear{??}{git}{2018c}]%
        {github-netty}
 \bibinfo{year}{2018}\natexlab{c}.
\newblock \bibinfo{title}{Open-Source {Netty} Repository at {GitHub}}.
\newblock
\newblock
\newblock
\shownote{\url{https://github.com/netty/netty}.}


\bibitem[\protect\citeauthoryear{??}{git}{2018d}]%
        {github-finagle}
 \bibinfo{year}{2018}\natexlab{d}.
\newblock \bibinfo{title}{Open-Source {Twitter} {Finagle} Repository at
  {GitHub}}.
\newblock
\newblock
\newblock
\shownote{\url{https://github.com/twitter/finagle}.}


\bibitem[\protect\citeauthoryear{??}{rx}{2018}]%
        {rx}
 \bibinfo{year}{2018}\natexlab{}.
\newblock \bibinfo{title}{{ReactiveX} project}.
\newblock
\newblock
\newblock
\shownote{\url{http://reactivex.io/languages.html}.}


\bibitem[\protect\citeauthoryear{Akram, Sartor, McKinley, and Eeckhout}{Akram
  et~al\mbox{.}}{2018}]%
        {Akram:2018:WGC:3192366.3192392}
\bibfield{author}{\bibinfo{person}{Shoaib Akram}, \bibinfo{person}{Jennifer~B.
  Sartor}, \bibinfo{person}{Kathryn~S. McKinley}, {and} \bibinfo{person}{Lieven
  Eeckhout}.} \bibinfo{year}{2018}\natexlab{}.
\newblock \showarticletitle{{Write-rationing Garbage Collection for Hybrid
  Memories}}. In \bibinfo{booktitle}{\emph{PLDI}}. \bibinfo{pages}{62--77}.
\newblock


\bibitem[\protect\citeauthoryear{Basin, Bortnikov, Braginsky, Golan-Gueta,
  Hillel, Keidar, and Sulamy}{Basin et~al\mbox{.}}{2017}]%
        {Basin:2017:KKM:3018743.3018761}
\bibfield{author}{\bibinfo{person}{Dmitry Basin}, \bibinfo{person}{Edward
  Bortnikov}, \bibinfo{person}{Anastasia Braginsky}, \bibinfo{person}{Guy
  Golan-Gueta}, \bibinfo{person}{Eshcar Hillel}, \bibinfo{person}{Idit Keidar},
  {and} \bibinfo{person}{Moshe Sulamy}.} \bibinfo{year}{2017}\natexlab{}.
\newblock \showarticletitle{KiWi: A Key-Value Map for Scalable Real-Time
  Analytics}. In \bibinfo{booktitle}{\emph{Proceedings of the 22Nd ACM SIGPLAN
  Symposium on Principles and Practice of Parallel Programming}}
  \emph{(\bibinfo{series}{PPoPP '17})}. \bibinfo{publisher}{ACM},
  \bibinfo{address}{New York, NY, USA}, \bibinfo{pages}{357--369}.
\newblock
\showISBNx{978-1-4503-4493-7}
\urldef\tempurl%
\url{https://doi.org/10.1145/3018743.3018761}
\showDOI{\tempurl}


\bibitem[\protect\citeauthoryear{Biswas, Cao, Zhang, Bond, and Wood}{Biswas
  et~al\mbox{.}}{2017}]%
        {Biswas:2017:LDR:3033019.3033020}
\bibfield{author}{\bibinfo{person}{Swarnendu Biswas}, \bibinfo{person}{Man
  Cao}, \bibinfo{person}{Minjia Zhang}, \bibinfo{person}{Michael~D. Bond},
  {and} \bibinfo{person}{Benjamin~P. Wood}.} \bibinfo{year}{2017}\natexlab{}.
\newblock \showarticletitle{{Lightweight Data Race Detection for Production
  Runs}}. In \bibinfo{booktitle}{\emph{CC}}. \bibinfo{pages}{11--21}.
\newblock


\bibitem[\protect\citeauthoryear{Blackburn, Garner, Hoffmann, Khang, McKinley,
  Bentzur, Diwan, Feinberg, Frampton, Guyer, Hirzel, Hosking, Jump, Lee, Moss,
  Phansalkar, Stefanovi\'{c}, VanDrunen, von Dincklage, and
  Wiedermann}{Blackburn et~al\mbox{.}}{2006}]%
        {Blackburn:2006:DBJ:1167515.1167488}
\bibfield{author}{\bibinfo{person}{Stephen~M. Blackburn},
  \bibinfo{person}{Robin Garner}, \bibinfo{person}{Chris Hoffmann},
  \bibinfo{person}{Asjad~M. Khang}, \bibinfo{person}{Kathryn~S. McKinley},
  \bibinfo{person}{Rotem Bentzur}, \bibinfo{person}{Amer Diwan},
  \bibinfo{person}{Daniel Feinberg}, \bibinfo{person}{Daniel Frampton},
  \bibinfo{person}{Samuel~Z. Guyer}, \bibinfo{person}{Martin Hirzel},
  \bibinfo{person}{Antony Hosking}, \bibinfo{person}{Maria Jump},
  \bibinfo{person}{Han Lee}, \bibinfo{person}{J.~Eliot~B. Moss},
  \bibinfo{person}{Aashish Phansalkar}, \bibinfo{person}{Darko Stefanovi\'{c}},
  \bibinfo{person}{Thomas VanDrunen}, \bibinfo{person}{Daniel von Dincklage},
  {and} \bibinfo{person}{Ben Wiedermann}.} \bibinfo{year}{2006}\natexlab{}.
\newblock \showarticletitle{{The DaCapo Benchmarks: Java Benchmarking
  Development and Analysis}}.
\newblock \bibinfo{journal}{\emph{SIGPLAN Not.}} \bibinfo{volume}{41},
  \bibinfo{number}{10} (\bibinfo{date}{Oct.} \bibinfo{year}{2006}),
  \bibinfo{pages}{169--190}.
\newblock


\bibitem[\protect\citeauthoryear{Bronson, Boner, Korland, Prokopec, Sankar,
  Spiewak, and Veentjer}{Bronson et~al\mbox{.}}{2011}]%
        {scala-stm-expert-group}
\bibfield{author}{\bibinfo{person}{Nathan Bronson}, \bibinfo{person}{Jonas
  Boner}, \bibinfo{person}{Guy Korland}, \bibinfo{person}{Aleksandar Prokopec},
  \bibinfo{person}{Krishna Sankar}, \bibinfo{person}{Daniel Spiewak}, {and}
  \bibinfo{person}{Peter Veentjer}.} \bibinfo{year}{2011}\natexlab{}.
\newblock \bibinfo{title}{Scala{S}{T}{M} Expert Group}.
\newblock
\newblock
\newblock
\shownote{\url{https://nbronson.github.io/scala-stm/expert\_group.html}.}


\bibitem[\protect\citeauthoryear{Bronson, Casper, Chafi, and Olukotun}{Bronson
  et~al\mbox{.}}{2010a}]%
        {bronsonavl}
\bibfield{author}{\bibinfo{person}{Nathan~G. Bronson}, \bibinfo{person}{Jared
  Casper}, \bibinfo{person}{Hassan Chafi}, {and} \bibinfo{person}{Kunle
  Olukotun}.} \bibinfo{year}{2010}\natexlab{a}.
\newblock \showarticletitle{A Practical Concurrent Binary Search Tree}.
\newblock \bibinfo{journal}{\emph{SIGPLAN Not.}} \bibinfo{volume}{45},
  \bibinfo{number}{5} (\bibinfo{date}{Jan.} \bibinfo{year}{2010}),
  \bibinfo{pages}{257--268}.
\newblock
\showISSN{0362-1340}
\urldef\tempurl%
\url{https://doi.org/10.1145/1837853.1693488}
\showDOI{\tempurl}


\bibitem[\protect\citeauthoryear{Bronson, Chafi, and Olukotun}{Bronson
  et~al\mbox{.}}{2010b}]%
        {Bronson10ccstm:a}
\bibfield{author}{\bibinfo{person}{Nathan~G. Bronson}, \bibinfo{person}{Hassan
  Chafi}, {and} \bibinfo{person}{Kunle Olukotun}.}
  \bibinfo{year}{2010}\natexlab{b}.
\newblock \showarticletitle{{CCSTM: A library-based STM for Scala}}. In
  \bibinfo{booktitle}{\emph{The First Annual Scala Workshop at Scala Days}}.
\newblock


\bibitem[\protect\citeauthoryear{Bruno and Ferreira}{Bruno and
  Ferreira}{2017}]%
        {Bruno:2017:PAP:3135974.3135986}
\bibfield{author}{\bibinfo{person}{Rodrigo Bruno} {and} \bibinfo{person}{Paulo
  Ferreira}.} \bibinfo{year}{2017}\natexlab{}.
\newblock \showarticletitle{{POLM2: Automatic Profiling for Object
  Lifetime-aware Memory Management for Hotspot Big Data Applications}}. In
  \bibinfo{booktitle}{\emph{Middleware}}. \bibinfo{pages}{147--160}.
\newblock


\bibitem[\protect\citeauthoryear{Calciu, Sen, Balakrishnan, and
  Aguilera}{Calciu et~al\mbox{.}}{2017}]%
        {Calciu:2017:BCD:3037697.3037721}
\bibfield{author}{\bibinfo{person}{Irina Calciu}, \bibinfo{person}{Siddhartha
  Sen}, \bibinfo{person}{Mahesh Balakrishnan}, {and} \bibinfo{person}{Marcos~K.
  Aguilera}.} \bibinfo{year}{2017}\natexlab{}.
\newblock \showarticletitle{Black-box Concurrent Data Structures for NUMA
  Architectures}. In \bibinfo{booktitle}{\emph{Proceedings of the Twenty-Second
  International Conference on Architectural Support for Programming Languages
  and Operating Systems}} \emph{(\bibinfo{series}{ASPLOS '17})}.
  \bibinfo{publisher}{ACM}, \bibinfo{address}{New York, NY, USA},
  \bibinfo{pages}{207--221}.
\newblock
\showISBNx{978-1-4503-4465-4}
\urldef\tempurl%
\url{https://doi.org/10.1145/3037697.3037721}
\showDOI{\tempurl}


\bibitem[\protect\citeauthoryear{Click}{Click}{2007}]%
        {cliffclick}
\bibfield{author}{\bibinfo{person}{Cliff Click}.}
  \bibinfo{year}{2007}\natexlab{}.
\newblock \bibinfo{title}{Towards a Scalable Non-Blocking Coding Style}.
\newblock
\newblock
\urldef\tempurl%
\url{http://www.azulsystems.com/events/javaone\_2007/2007\_LockFreeHash.pdf}
\showURL{%
\tempurl}


\bibitem[\protect\citeauthoryear{Databricks}{Databricks}{2018}]%
        {spark-perf}
\bibfield{author}{\bibinfo{person}{Databricks}.}
  \bibinfo{year}{2018}\natexlab{}.
\newblock \bibinfo{title}{Spark Performance Tests}.
\newblock
  \bibinfo{howpublished}{\url{https://github.com/databricks/spark-perf}}.
\newblock


\bibitem[\protect\citeauthoryear{Duboscq}{Duboscq}{2016}]%
        {duboscq-thesis}
\bibfield{author}{\bibinfo{person}{Gilles Duboscq}.}
  \bibinfo{year}{2016}\natexlab{}.
\newblock \emph{\bibinfo{title}{{Combining Speculative Optimizations with
  Flexible Scheduling of Side-effects}}}.
\newblock \bibinfo{thesistype}{Ph.D. Dissertation}. \bibinfo{school}{Johannes
  Kepler University, Linz}.
\newblock


\bibitem[\protect\citeauthoryear{Duboscq, W\"{u}rthinger, Stadler, Wimmer,
  Simon, and M\"{o}ssenb\"{o}ck}{Duboscq et~al\mbox{.}}{2013}]%
        {Duboscq:2013:IRS:2542142.2542143}
\bibfield{author}{\bibinfo{person}{Gilles Duboscq}, \bibinfo{person}{Thomas
  W\"{u}rthinger}, \bibinfo{person}{Lukas Stadler}, \bibinfo{person}{Christian
  Wimmer}, \bibinfo{person}{Doug Simon}, {and} \bibinfo{person}{Hanspeter
  M\"{o}ssenb\"{o}ck}.} \bibinfo{year}{2013}\natexlab{}.
\newblock \showarticletitle{{An Intermediate Representation for Speculative
  Optimizations in a Dynamic Compiler}}. In \bibinfo{booktitle}{\emph{VMIL}}.
  \bibinfo{pages}{1--10}.
\newblock


\bibitem[\protect\citeauthoryear{Duigou}{Duigou}{2011}]%
        {jep107}
\bibfield{author}{\bibinfo{person}{Michael Duigou}.}
  \bibinfo{year}{2011}\natexlab{}.
\newblock \bibinfo{title}{{Java} {Enhancement} {Proposal} 107: {Bulk} {Data}
  {Operations} for {Collections}}.
\newblock
\newblock
\newblock
\shownote{\url{http://openjdk.java.net/jeps/107}.}


\bibitem[\protect\citeauthoryear{Eisl, Grimmer, Simon, W\"{u}rthinger, and
  M\"{o}ssenb\"{o}ck}{Eisl et~al\mbox{.}}{2016}]%
        {Eisl:2016:TRA:2972206.2972211}
\bibfield{author}{\bibinfo{person}{Josef Eisl}, \bibinfo{person}{Matthias
  Grimmer}, \bibinfo{person}{Doug Simon}, \bibinfo{person}{Thomas
  W\"{u}rthinger}, {and} \bibinfo{person}{Hanspeter M\"{o}ssenb\"{o}ck}.}
  \bibinfo{year}{2016}\natexlab{}.
\newblock \showarticletitle{{Trace-based Register Allocation in a JIT
  Compiler}}. In \bibinfo{booktitle}{\emph{PPPJ}}.
  \bibinfo{pages}{14:1--14:11}.
\newblock


\bibitem[\protect\citeauthoryear{Ferdman, Adileh, Kocberber, Volos, Alisafaee,
  Jevdjic, Kaynak, Popescu, Ailamaki, and Falsafi}{Ferdman
  et~al\mbox{.}}{2012}]%
        {Ferdman:2012:CCS:2248487.2150982}
\bibfield{author}{\bibinfo{person}{Michael Ferdman}, \bibinfo{person}{Almutaz
  Adileh}, \bibinfo{person}{Onur Kocberber}, \bibinfo{person}{Stavros Volos},
  \bibinfo{person}{Mohammad Alisafaee}, \bibinfo{person}{Djordje Jevdjic},
  \bibinfo{person}{Cansu Kaynak}, \bibinfo{person}{Adrian~Daniel Popescu},
  \bibinfo{person}{Anastasia Ailamaki}, {and} \bibinfo{person}{Babak Falsafi}.}
  \bibinfo{year}{2012}\natexlab{}.
\newblock \showarticletitle{{Clearing the Clouds: A Study of Emerging Scale-out
  Workloads on Modern Hardware}}.
\newblock \bibinfo{journal}{\emph{SIGPLAN Not.}} \bibinfo{volume}{47},
  \bibinfo{number}{4} (\bibinfo{date}{March} \bibinfo{year}{2012}),
  \bibinfo{pages}{37--48}.
\newblock


\bibitem[\protect\citeauthoryear{Grech, Fourtounis, Francalanza, and
  Smaragdakis}{Grech et~al\mbox{.}}{2018}]%
        {Grech:2018:SHU:3213846.3213860}
\bibfield{author}{\bibinfo{person}{Neville Grech}, \bibinfo{person}{George
  Fourtounis}, \bibinfo{person}{Adrian Francalanza}, {and}
  \bibinfo{person}{Yannis Smaragdakis}.} \bibinfo{year}{2018}\natexlab{}.
\newblock \showarticletitle{{Shooting from the Heap: Ultra-scalable Static
  Analysis with Heap Snapshots}}. In \bibinfo{booktitle}{\emph{ISSTA}}.
  \bibinfo{pages}{198--208}.
\newblock


\bibitem[\protect\citeauthoryear{Guerraoui, Kapalka, and Vitek}{Guerraoui
  et~al\mbox{.}}{2007}]%
        {Guerraoui:2007:SBS:1272996.1273029}
\bibfield{author}{\bibinfo{person}{Rachid Guerraoui}, \bibinfo{person}{Michal
  Kapalka}, {and} \bibinfo{person}{Jan Vitek}.}
  \bibinfo{year}{2007}\natexlab{}.
\newblock \showarticletitle{{STMBench7: A Benchmark for Software Transactional
  Memory}}. In \bibinfo{booktitle}{\emph{EuroSys}}. \bibinfo{pages}{315--324}.
\newblock


\bibitem[\protect\citeauthoryear{Haller and Odersky}{Haller and
  Odersky}{2007}]%
        {Haller:2007:AUT:1764606.1764620}
\bibfield{author}{\bibinfo{person}{Philipp Haller} {and}
  \bibinfo{person}{Martin Odersky}.} \bibinfo{year}{2007}\natexlab{}.
\newblock \showarticletitle{{Actors That Unify Threads and Events}}. In
  \bibinfo{booktitle}{\emph{COORDINATION}}. \bibinfo{pages}{171--190}.
\newblock


\bibitem[\protect\citeauthoryear{Haller, Prokopec, Miller, Klang, Kuhn, and
  Jovanovic}{Haller et~al\mbox{.}}{2012}]%
        {SIP14}
\bibfield{author}{\bibinfo{person}{Philipp Haller}, \bibinfo{person}{Aleksandar
  Prokopec}, \bibinfo{person}{Heather Miller}, \bibinfo{person}{Viktor Klang},
  \bibinfo{person}{Roland Kuhn}, {and} \bibinfo{person}{Vojin Jovanovic}.}
  \bibinfo{year}{2012}\natexlab{}.
\newblock \showarticletitle{Scala Improvement Proposal: Futures and Promises
  ({SIP}-14)}.
\newblock
\urldef\tempurl%
\url{http://docs.scala-lang.org/sips/pending/futures-promises.html}
\showURL{%
\tempurl}


\bibitem[\protect\citeauthoryear{Hofer, Gnedt, Sch\"{o}rgenhumer, and
  M\"{o}ssenb\"{o}ck}{Hofer et~al\mbox{.}}{2016}]%
        {Hofer:2016:ETV:2851553.2851559}
\bibfield{author}{\bibinfo{person}{Peter Hofer}, \bibinfo{person}{David Gnedt},
  \bibinfo{person}{Andreas Sch\"{o}rgenhumer}, {and} \bibinfo{person}{Hanspeter
  M\"{o}ssenb\"{o}ck}.} \bibinfo{year}{2016}\natexlab{}.
\newblock \showarticletitle{{Efficient Tracing and Versatile Analysis of Lock
  Contention in Java Applications on the Virtual Machine Level}}. In
  \bibinfo{booktitle}{\emph{ICPE}}. \bibinfo{pages}{263--274}.
\newblock


\bibitem[\protect\citeauthoryear{{Intel}}{{Intel}}{2018a}]%
        {hyperThreading}
\bibfield{author}{\bibinfo{person}{{Intel}}.} \bibinfo{year}{2018}\natexlab{a}.
\newblock \bibinfo{title}{{Hyper-Threading Technology}}.
\newblock
  \bibinfo{howpublished}{\url{https://www.intel.com/content/www/us/en/architecture-and-technology/hyper-threading/hyper-threading-technology.html}}.
\newblock


\bibitem[\protect\citeauthoryear{{Intel}}{{Intel}}{2018b}]%
        {turboboost}
\bibfield{author}{\bibinfo{person}{{Intel}}.} \bibinfo{year}{2018}\natexlab{b}.
\newblock \bibinfo{title}{{Turbo Boost Technology 2.0}}.
\newblock
  \bibinfo{howpublished}{\url{https://www.intel.com/content/www/us/en/architecture-and-technology/turbo-boost/turbo-boost-technology.html}}.
\newblock


\bibitem[\protect\citeauthoryear{Lea}{Lea}{2000}]%
        {Lea:2000:JFF:337449.337465}
\bibfield{author}{\bibinfo{person}{Doug Lea}.} \bibinfo{year}{2000}\natexlab{}.
\newblock \showarticletitle{{A Java Fork/Join Framework}}. In
  \bibinfo{booktitle}{\emph{JAVA}}. \bibinfo{pages}{36--43}.
\newblock


\bibitem[\protect\citeauthoryear{Lea}{Lea}{2014}]%
        {dougleahome}
\bibfield{author}{\bibinfo{person}{Doug Lea}.} \bibinfo{year}{2014}\natexlab{}.
\newblock \bibinfo{title}{{D}oug {L}ea's Workstation}.
\newblock
\newblock
\urldef\tempurl%
\url{http://g.oswego.edu/}
\showURL{%
\tempurl}


\bibitem[\protect\citeauthoryear{Lengauer, Bitto, M\"{o}ssenb\"{o}ck, and
  Weninger}{Lengauer et~al\mbox{.}}{2017}]%
        {Lengauer:2017:CJB:3030207.3030211}
\bibfield{author}{\bibinfo{person}{Philipp Lengauer}, \bibinfo{person}{Verena
  Bitto}, \bibinfo{person}{Hanspeter M\"{o}ssenb\"{o}ck}, {and}
  \bibinfo{person}{Markus Weninger}.} \bibinfo{year}{2017}\natexlab{}.
\newblock \showarticletitle{{A Comprehensive {Java} Benchmark Study on Memory
  and Garbage Collection Behavior of {DaCapo}, {DaCapo} {Scala}, and
  {SPECjvm2008}}}. In \bibinfo{booktitle}{\emph{ICPE}}. \bibinfo{pages}{3--14}.
\newblock


\bibitem[\protect\citeauthoryear{Leopoldseder, Stadler, W\"{u}rthinger, Eisl,
  Simon, and M\"{o}ssenb\"{o}ck}{Leopoldseder et~al\mbox{.}}{2018}]%
        {Leopoldseder:2018:DDS:3179541.3168811}
\bibfield{author}{\bibinfo{person}{David Leopoldseder}, \bibinfo{person}{Lukas
  Stadler}, \bibinfo{person}{Thomas W\"{u}rthinger}, \bibinfo{person}{Josef
  Eisl}, \bibinfo{person}{Doug Simon}, {and} \bibinfo{person}{Hanspeter
  M\"{o}ssenb\"{o}ck}.} \bibinfo{year}{2018}\natexlab{}.
\newblock \showarticletitle{{Dominance-based Duplication Simulation (DBDS):
  Code Duplication to Enable Compiler Optimizations}}. In
  \bibinfo{booktitle}{\emph{CGO}}. \bibinfo{pages}{126--137}.
\newblock


\bibitem[\protect\citeauthoryear{Liu and Huang}{Liu and Huang}{2018}]%
        {Liu:2018:DFC:3192366.3192390}
\bibfield{author}{\bibinfo{person}{Bozhen Liu} {and} \bibinfo{person}{Jeff
  Huang}.} \bibinfo{year}{2018}\natexlab{}.
\newblock \showarticletitle{{D4: Fast Concurrency Debugging with Parallel
  Differential Analysis}}. In \bibinfo{booktitle}{\emph{PLDI}}.
  \bibinfo{pages}{359--373}.
\newblock


\bibitem[\protect\citeauthoryear{Lu, Cox, and Zwaenepoel}{Lu
  et~al\mbox{.}}{2001}]%
        {Lu:2001:CER:379539.379568}
\bibfield{author}{\bibinfo{person}{Honghui Lu}, \bibinfo{person}{Alan~L. Cox},
  {and} \bibinfo{person}{Willy Zwaenepoel}.} \bibinfo{year}{2001}\natexlab{}.
\newblock \showarticletitle{Contention Elimination by Replication of Sequential
  Sections in Distributed Shared Memory Programs}. In
  \bibinfo{booktitle}{\emph{Proceedings of the Eighth ACM SIGPLAN Symposium on
  Principles and Practices of Parallel Programming}}
  \emph{(\bibinfo{series}{PPoPP '01})}. \bibinfo{publisher}{ACM},
  \bibinfo{address}{New York, NY, USA}, \bibinfo{pages}{53--61}.
\newblock
\showISBNx{1-58113-346-4}
\urldef\tempurl%
\url{https://doi.org/10.1145/379539.379568}
\showDOI{\tempurl}


\bibitem[\protect\citeauthoryear{Meng, Bradley, Yavuz, Sparks, Venkataraman,
  Liu, Freeman, Tsai, Amde, Owen, Xin, Xin, Franklin, Zadeh, Zaharia, and
  Talwalkar}{Meng et~al\mbox{.}}{2016}]%
        {JMLR:v17:15-237}
\bibfield{author}{\bibinfo{person}{Xiangrui Meng}, \bibinfo{person}{Joseph
  Bradley}, \bibinfo{person}{Burak Yavuz}, \bibinfo{person}{Evan Sparks},
  \bibinfo{person}{Shivaram Venkataraman}, \bibinfo{person}{Davies Liu},
  \bibinfo{person}{Jeremy Freeman}, \bibinfo{person}{DB Tsai},
  \bibinfo{person}{Manish Amde}, \bibinfo{person}{Sean Owen},
  \bibinfo{person}{Doris Xin}, \bibinfo{person}{Reynold Xin},
  \bibinfo{person}{Michael~J. Franklin}, \bibinfo{person}{Reza Zadeh},
  \bibinfo{person}{Matei Zaharia}, {and} \bibinfo{person}{Ameet Talwalkar}.}
  \bibinfo{year}{2016}\natexlab{}.
\newblock \showarticletitle{{MLlib: Machine Learning in Apache Spark}}.
\newblock \bibinfo{journal}{\emph{Journal of Machine Learning Research}}
  \bibinfo{volume}{17}, \bibinfo{number}{34} (\bibinfo{year}{2016}),
  \bibinfo{pages}{1--7}.
\newblock


\bibitem[\protect\citeauthoryear{Nguyen, Fang, Xu, Demsky, Lu, Alamian, and
  Mutlu}{Nguyen et~al\mbox{.}}{2016}]%
        {Nguyen:2016:YHB:3026877.3026905}
\bibfield{author}{\bibinfo{person}{Khanh Nguyen}, \bibinfo{person}{Lu Fang},
  \bibinfo{person}{Guoqing Xu}, \bibinfo{person}{Brian Demsky},
  \bibinfo{person}{Shan Lu}, \bibinfo{person}{Sanazsadat Alamian}, {and}
  \bibinfo{person}{Onur Mutlu}.} \bibinfo{year}{2016}\natexlab{}.
\newblock \showarticletitle{{Yak: A High-performance Big-data-friendly Garbage
  Collector}}. In \bibinfo{booktitle}{\emph{OSDI}}. \bibinfo{pages}{349--365}.
\newblock


\bibitem[\protect\citeauthoryear{Odersky}{Odersky}{2011}]%
        {state-of-scala}
\bibfield{author}{\bibinfo{person}{Martin Odersky}.}
  \bibinfo{year}{2011}\natexlab{}.
\newblock \bibinfo{title}{State of {Scala}}.
\newblock
\newblock
\newblock
\shownote{\url{http://days2011.scala-lang.org/sites/days2011/files/01.\%20Martin\%20Odersky.pdf}.}


\bibitem[\protect\citeauthoryear{Oshman and Shavit}{Oshman and Shavit}{2013}]%
        {Oshman:2013:SLC:2484239.2484270}
\bibfield{author}{\bibinfo{person}{Rotem Oshman} {and} \bibinfo{person}{Nir
  Shavit}.} \bibinfo{year}{2013}\natexlab{}.
\newblock \showarticletitle{The SkipTrie: Low-depth Concurrent Search Without
  Rebalancing}. In \bibinfo{booktitle}{\emph{Proceedings of the 2013 ACM
  Symposium on Principles of Distributed Computing}}
  \emph{(\bibinfo{series}{PODC '13})}. \bibinfo{publisher}{ACM},
  \bibinfo{address}{New York, NY, USA}, \bibinfo{pages}{23--32}.
\newblock
\showISBNx{978-1-4503-2065-8}
\urldef\tempurl%
\url{https://doi.org/10.1145/2484239.2484270}
\showDOI{\tempurl}


\bibitem[\protect\citeauthoryear{Paleczny, Vick, and Click}{Paleczny
  et~al\mbox{.}}{2001}]%
        {Paleczny:2001:JHT:1267847.1267848}
\bibfield{author}{\bibinfo{person}{Michael Paleczny},
  \bibinfo{person}{Christopher Vick}, {and} \bibinfo{person}{Cliff Click}.}
  \bibinfo{year}{2001}\natexlab{}.
\newblock \showarticletitle{The {Java} {HotspotTM} Server Compiler}. In
  \bibinfo{booktitle}{\emph{JVM}}.
\newblock


\bibitem[\protect\citeauthoryear{Paumard}{Paumard}{2018}]%
        {scrabble-benchmark}
\bibfield{author}{\bibinfo{person}{Jos{\'e} Paumard}.}
  \bibinfo{year}{2018}\natexlab{}.
\newblock \bibinfo{title}{JDK8 Stream/Rx Comparison}.
\newblock
  \bibinfo{howpublished}{\url{https://github.com/JosePaumard/jdk8-stream-rx-comparison}}.
\newblock


\bibitem[\protect\citeauthoryear{Prokopec}{Prokopec}{2015}]%
        {Prokopec:2015:SLQ:2774975.2774976}
\bibfield{author}{\bibinfo{person}{Aleksandar Prokopec}.}
  \bibinfo{year}{2015}\natexlab{}.
\newblock \showarticletitle{Snap{Q}ueue: Lock-free Queue with Constant Time
  Snapshots}. In \bibinfo{booktitle}{\emph{Proceedings of the 6th ACM SIGPLAN
  Symposium on Scala}} \emph{(\bibinfo{series}{SCALA 2015})}.
  \bibinfo{publisher}{ACM}, \bibinfo{address}{New York, NY, USA},
  \bibinfo{pages}{1--12}.
\newblock
\showISBNx{978-1-4503-3626-0}
\urldef\tempurl%
\url{https://doi.org/10.1145/2774975.2774976}
\showDOI{\tempurl}


\bibitem[\protect\citeauthoryear{Prokopec}{Prokopec}{2016}]%
        {Prokopec:2016:PSR:3001886.3001891}
\bibfield{author}{\bibinfo{person}{Aleksandar Prokopec}.}
  \bibinfo{year}{2016}\natexlab{}.
\newblock \showarticletitle{{Pluggable Scheduling for the Reactor Programming
  Model}}. In \bibinfo{booktitle}{\emph{AGERE!}} \bibinfo{pages}{41--50}.
\newblock


\bibitem[\protect\citeauthoryear{Prokopec}{Prokopec}{2017a}]%
        {Prokopec2017}
\bibfield{author}{\bibinfo{person}{Aleksandar Prokopec}.}
  \bibinfo{year}{2017}\natexlab{a}.
\newblock \bibinfo{booktitle}{\emph{Accelerating by Idling: How Speculative
  Delays Improve Performance of Message-Oriented Systems}}.
\newblock \bibinfo{publisher}{Springer International Publishing},
  \bibinfo{address}{Cham}, \bibinfo{pages}{177--191}.
\newblock
\showISBNx{978-3-319-64203-1}
\urldef\tempurl%
\url{https://doi.org/10.1007/978-3-319-64203-1_13}
\showDOI{\tempurl}


\bibitem[\protect\citeauthoryear{Prokopec}{Prokopec}{2017b}]%
        {techreport-17-prokopec}
\bibfield{author}{\bibinfo{person}{Aleksandar Prokopec}.}
  \bibinfo{year}{2017}\natexlab{b}.
\newblock \showarticletitle{{Analysis of Concurrent Lock-Free Hash Tries with
  Constant-Time Operations}}.
\newblock \bibinfo{journal}{\emph{ArXiv e-prints}} (\bibinfo{date}{Dec.}
  \bibinfo{year}{2017}).
\newblock
\showeprint[arxiv]{cs.DS/1712.09636}


\bibitem[\protect\citeauthoryear{Prokopec}{Prokopec}{2017c}]%
        {Prokopec:2017:EBB:3133850.3133865}
\bibfield{author}{\bibinfo{person}{Aleksandar Prokopec}.}
  \bibinfo{year}{2017}\natexlab{c}.
\newblock \showarticletitle{Encoding the Building Blocks of Communication}. In
  \bibinfo{booktitle}{\emph{Proceedings of the 2017 ACM SIGPLAN International
  Symposium on New Ideas, New Paradigms, and Reflections on Programming and
  Software}} \emph{(\bibinfo{series}{Onward! 2017})}. \bibinfo{publisher}{ACM},
  \bibinfo{address}{New York, NY, USA}, \bibinfo{pages}{104--118}.
\newblock
\showISBNx{978-1-4503-5530-8}
\urldef\tempurl%
\url{https://doi.org/10.1145/3133850.3133865}
\showDOI{\tempurl}


\bibitem[\protect\citeauthoryear{Prokopec}{Prokopec}{2018a}]%
        {Prokopec:2018:CCL:3178487.3178498}
\bibfield{author}{\bibinfo{person}{Aleksandar Prokopec}.}
  \bibinfo{year}{2018}\natexlab{a}.
\newblock \showarticletitle{Cache-tries: Concurrent Lock-free Hash Tries with
  Constant-time Operations}. In \bibinfo{booktitle}{\emph{Proceedings of the
  23rd ACM SIGPLAN Symposium on Principles and Practice of Parallel
  Programming}} \emph{(\bibinfo{series}{PPoPP '18})}. \bibinfo{publisher}{ACM},
  \bibinfo{address}{New York, NY, USA}, \bibinfo{pages}{137--151}.
\newblock
\showISBNx{978-1-4503-4982-6}
\urldef\tempurl%
\url{https://doi.org/10.1145/3178487.3178498}
\showDOI{\tempurl}


\bibitem[\protect\citeauthoryear{Prokopec}{Prokopec}{2018b}]%
        {Prokopec2018}
\bibfield{author}{\bibinfo{person}{Aleksandar Prokopec}.}
  \bibinfo{year}{2018}\natexlab{b}.
\newblock \bibinfo{booktitle}{\emph{Efficient Lock-Free Removing and Compaction
  for the Cache-Trie Data Structure}}.
\newblock \bibinfo{publisher}{Springer International Publishing},
  \bibinfo{address}{Cham}.
\newblock


\bibitem[\protect\citeauthoryear{Prokopec}{Prokopec}{2018c}]%
        {cache-trie-remove-dataset}
\bibfield{author}{\bibinfo{person}{Aleksandar Prokopec}.}
  \bibinfo{year}{2018}\natexlab{c}.
\newblock \bibinfo{title}{Efficient Lock-Free Removing and Compaction for the
  Cache-Trie Data Structure}.
\newblock \bibinfo{howpublished}{https://doi.org/10.6084/m9.figshare.6369134}.
\newblock


\bibitem[\protect\citeauthoryear{Prokopec, Bagwell, and Odersky}{Prokopec
  et~al\mbox{.}}{2011a}]%
        {EPFL-REPORT-166908}
\bibfield{author}{\bibinfo{person}{Aleksandar Prokopec}, \bibinfo{person}{Phil
  Bagwell}, {and} \bibinfo{person}{Martin Odersky}.}
  \bibinfo{year}{2011}\natexlab{a}.
\newblock \bibinfo{booktitle}{\emph{Cache-{A}ware {L}ock-{F}ree {C}oncurrent
  {H}ash {T}ries}}.
\newblock \bibinfo{type}{{T}echnical {R}eport}.
\newblock


\bibitem[\protect\citeauthoryear{Prokopec, Bagwell, and Odersky}{Prokopec
  et~al\mbox{.}}{2011b}]%
        {Prokopec2011}
\bibfield{author}{\bibinfo{person}{Aleksandar Prokopec}, \bibinfo{person}{Phil
  Bagwell}, {and} \bibinfo{person}{Martin Odersky}.}
  \bibinfo{year}{2011}\natexlab{b}.
\newblock \bibinfo{booktitle}{\emph{Lock-Free Resizeable Concurrent Tries}}.
\newblock \bibinfo{publisher}{Springer Berlin Heidelberg},
  \bibinfo{address}{Berlin, Heidelberg}, \bibinfo{pages}{156--170}.
\newblock
\showISBNx{978-3-642-36036-7}
\urldef\tempurl%
\url{https://doi.org/10.1007/978-3-642-36036-7_11}
\showDOI{\tempurl}


\bibitem[\protect\citeauthoryear{Prokopec, Bagwell, Rompf, and
  Odersky}{Prokopec et~al\mbox{.}}{2011c}]%
        {prokopec11}
\bibfield{author}{\bibinfo{person}{Aleksandar Prokopec}, \bibinfo{person}{Phil
  Bagwell}, \bibinfo{person}{Tiark Rompf}, {and} \bibinfo{person}{Martin
  Odersky}.} \bibinfo{year}{2011}\natexlab{c}.
\newblock \showarticletitle{{A Generic Parallel Collection Framework}}. In
  \bibinfo{booktitle}{\emph{Euro-Par}}. \bibinfo{pages}{136--147}.
\newblock


\bibitem[\protect\citeauthoryear{Prokopec, Bronson, Bagwell, and
  Odersky}{Prokopec et~al\mbox{.}}{2012a}]%
        {prokopec12ctries}
\bibfield{author}{\bibinfo{person}{Aleksandar Prokopec},
  \bibinfo{person}{Nathan~Grasso Bronson}, \bibinfo{person}{Phil Bagwell},
  {and} \bibinfo{person}{Martin Odersky}.} \bibinfo{year}{2012}\natexlab{a}.
\newblock \showarticletitle{Concurrent Tries with Efficient Non-blocking
  Snapshots}. In \bibinfo{booktitle}{\emph{Proceedings of the 17th ACM SIGPLAN
  Symposium on Principles and Practice of Parallel Programming}}
  \emph{(\bibinfo{series}{PPoPP '12})}. \bibinfo{publisher}{ACM},
  \bibinfo{address}{New York, NY, USA}, \bibinfo{pages}{151--160}.
\newblock
\showISBNx{978-1-4503-1160-1}
\urldef\tempurl%
\url{https://doi.org/10.1145/2145816.2145836}
\showDOI{\tempurl}


\bibitem[\protect\citeauthoryear{Prokopec, Duboscq, Leopoldseder, and
  W\"{u}rthinger}{Prokopec et~al\mbox{.}}{2019a}]%
        {Prokopec:2019:OII:3314872.3314893}
\bibfield{author}{\bibinfo{person}{Aleksandar Prokopec},
  \bibinfo{person}{Gilles Duboscq}, \bibinfo{person}{David Leopoldseder}, {and}
  \bibinfo{person}{Thomas W\"{u}rthinger}.} \bibinfo{year}{2019}\natexlab{a}.
\newblock \showarticletitle{An Optimization-driven Incremental Inline
  Substitution Algorithm for Just-in-time Compilers}. In
  \bibinfo{booktitle}{\emph{Proceedings of the 2019 IEEE/ACM International
  Symposium on Code Generation and Optimization}} \emph{(\bibinfo{series}{CGO
  2019})}. \bibinfo{publisher}{IEEE Press}, \bibinfo{address}{Piscataway, NJ,
  USA}, \bibinfo{pages}{164--179}.
\newblock
\showISBNx{978-1-7281-1436-1}
\urldef\tempurl%
\url{http://dl.acm.org/citation.cfm?id=3314872.3314893}
\showURL{%
\tempurl}


\bibitem[\protect\citeauthoryear{Prokopec, Haller, and Odersky}{Prokopec
  et~al\mbox{.}}{2014a}]%
        {Prokopec:2014:CAM:2637647.2637656}
\bibfield{author}{\bibinfo{person}{Aleksandar Prokopec},
  \bibinfo{person}{Philipp Haller}, {and} \bibinfo{person}{Martin Odersky}.}
  \bibinfo{year}{2014}\natexlab{a}.
\newblock \showarticletitle{Containers and Aggregates, Mutators and Isolates
  for Reactive Programming}. In \bibinfo{booktitle}{\emph{Proceedings of the
  Fifth Annual Scala Workshop}} \emph{(\bibinfo{series}{SCALA '14})}.
  \bibinfo{publisher}{ACM}, \bibinfo{address}{New York, NY, USA},
  \bibinfo{pages}{51--61}.
\newblock
\showISBNx{978-1-4503-2868-5}
\urldef\tempurl%
\url{https://doi.org/10.1145/2637647.2637656}
\showDOI{\tempurl}


\bibitem[\protect\citeauthoryear{Prokopec, Leopoldseder, Duboscq, and
  W\"{u}rthinger}{Prokopec et~al\mbox{.}}{2017}]%
        {Prokopec:2017:MCO:3136000.3136002}
\bibfield{author}{\bibinfo{person}{Aleksandar Prokopec}, \bibinfo{person}{David
  Leopoldseder}, \bibinfo{person}{Gilles Duboscq}, {and}
  \bibinfo{person}{Thomas W\"{u}rthinger}.} \bibinfo{year}{2017}\natexlab{}.
\newblock \showarticletitle{{Making Collection Operations Optimal with
  Aggressive JIT Compilation}}. In \bibinfo{booktitle}{\emph{SCALA}}.
  \bibinfo{pages}{29--40}.
\newblock


\bibitem[\protect\citeauthoryear{Prokopec and Liu}{Prokopec and Liu}{2018a}]%
        {coroutines-tech-report}
\bibfield{author}{\bibinfo{person}{Aleksandar Prokopec} {and}
  \bibinfo{person}{Fengyun Liu}.} \bibinfo{year}{2018}\natexlab{a}.
\newblock \showarticletitle{On the Soundness of Coroutines with Snapshots}.
\newblock \bibinfo{journal}{\emph{CoRR}}  \bibinfo{volume}{abs/1806.01405}
  (\bibinfo{year}{2018}).
\newblock
\showeprint[arxiv]{1806.01405}
\urldef\tempurl%
\url{https://arxiv.org/abs/1806.01405}
\showURL{%
\tempurl}


\bibitem[\protect\citeauthoryear{Prokopec and Liu}{Prokopec and Liu}{2018b}]%
        {DBLP:conf/ecoop/ProkopecL18}
\bibfield{author}{\bibinfo{person}{Aleksandar Prokopec} {and}
  \bibinfo{person}{Fengyun Liu}.} \bibinfo{year}{2018}\natexlab{b}.
\newblock \showarticletitle{Theory and Practice of Coroutines with Snapshots}.
  In \bibinfo{booktitle}{\emph{32nd European Conference on Object-Oriented
  Programming, {ECOOP} 2018, July 16-21, 2018, Amsterdam, The Netherlands}}.
  \bibinfo{pages}{3:1--3:32}.
\newblock
\urldef\tempurl%
\url{https://doi.org/10.4230/LIPIcs.ECOOP.2018.3}
\showDOI{\tempurl}


\bibitem[\protect\citeauthoryear{Prokopec, Miller, Schlatter, Haller, and
  Odersky}{Prokopec et~al\mbox{.}}{2012b}]%
        {prokopec12flowpools}
\bibfield{author}{\bibinfo{person}{Aleksandar Prokopec},
  \bibinfo{person}{Heather Miller}, \bibinfo{person}{Tobias Schlatter},
  \bibinfo{person}{Philipp Haller}, {and} \bibinfo{person}{Martin Odersky}.}
  \bibinfo{year}{2012}\natexlab{b}.
\newblock \showarticletitle{FlowPools: A Lock-Free Deterministic Concurrent
  Dataflow Abstraction}. In \bibinfo{booktitle}{\emph{LCPC}}.
  \bibinfo{pages}{158--173}.
\newblock


\bibitem[\protect\citeauthoryear{Prokopec and Odersky}{Prokopec and
  Odersky}{2014}]%
        {10.1007/978-3-319-09967-54}
\bibfield{author}{\bibinfo{person}{Aleksandar Prokopec} {and}
  \bibinfo{person}{Martin Odersky}.} \bibinfo{year}{2014}\natexlab{}.
\newblock \showarticletitle{Near Optimal Work-Stealing Tree Scheduler for
  Highly Irregular Data-Parallel Workloads}. In
  \bibinfo{booktitle}{\emph{Languages and Compilers for Parallel Computing}},
  \bibfield{editor}{\bibinfo{person}{C{\u{a}}lin Cascaval} {and}
  \bibinfo{person}{Pablo Montesinos}} (Eds.). \bibinfo{publisher}{Springer
  International Publishing}, \bibinfo{address}{Cham}, \bibinfo{pages}{55--86}.
\newblock
\showISBNx{978-3-319-09967-5}


\bibitem[\protect\citeauthoryear{Prokopec and Odersky}{Prokopec and
  Odersky}{2015}]%
        {Prokopec:2015:ICE:2814228.2814245}
\bibfield{author}{\bibinfo{person}{Aleksandar Prokopec} {and}
  \bibinfo{person}{Martin Odersky}.} \bibinfo{year}{2015}\natexlab{}.
\newblock \showarticletitle{{Isolates, Channels, and Event Streams for
  Composable Distributed Programming}}. In \bibinfo{booktitle}{\emph{Onward!}}
  \bibinfo{pages}{171--182}.
\newblock


\bibitem[\protect\citeauthoryear{Prokopec and Odersky}{Prokopec and
  Odersky}{2016}]%
        {Prokopec2016}
\bibfield{author}{\bibinfo{person}{Aleksandar Prokopec} {and}
  \bibinfo{person}{Martin Odersky}.} \bibinfo{year}{2016}\natexlab{}.
\newblock \bibinfo{booktitle}{\emph{Conc-Trees for Functional and Parallel
  Programming}}.
\newblock \bibinfo{publisher}{Springer International Publishing},
  \bibinfo{address}{Cham}, \bibinfo{pages}{254--268}.
\newblock
\showISBNx{978-3-319-29778-1}
\urldef\tempurl%
\url{https://doi.org/10.1007/978-3-319-29778-1\_16}
\showDOI{\tempurl}


\bibitem[\protect\citeauthoryear{Prokopec, Petrashko, and Odersky}{Prokopec
  et~al\mbox{.}}{2014b}]%
        {Prokopec:196627}
\bibfield{author}{\bibinfo{person}{Aleksandar Prokopec},
  \bibinfo{person}{Dmitry Petrashko}, {and} \bibinfo{person}{Martin Odersky}.}
  \bibinfo{year}{2014}\natexlab{b}.
\newblock \showarticletitle{On Lock-Free Work-stealing Iterators for Parallel
  Data Structures}.
\newblock  (\bibinfo{year}{2014}), \bibinfo{pages}{10}.
\newblock


\bibitem[\protect\citeauthoryear{Prokopec, Petrashko, and Odersky}{Prokopec
  et~al\mbox{.}}{2015}]%
        {7092728}
\bibfield{author}{\bibinfo{person}{A. Prokopec}, \bibinfo{person}{D.
  Petrashko}, {and} \bibinfo{person}{M. Odersky}.}
  \bibinfo{year}{2015}\natexlab{}.
\newblock \showarticletitle{Efficient Lock-Free Work-Stealing Iterators for
  Data-Parallel Collections}. In \bibinfo{booktitle}{\emph{2015 23rd Euromicro
  International Conference on Parallel, Distributed, and Network-Based
  Processing}}. \bibinfo{pages}{248--252}.
\newblock
\showISSN{1066-6192}
\urldef\tempurl%
\url{https://doi.org/10.1109/PDP.2015.65}
\showDOI{\tempurl}


\bibitem[\protect\citeauthoryear{Prokopec, Rosà, Leopoldseder, Duboscq, Tůma,
  Studener, Bulej, Zheng, Villazón, Simon, Würthinger, and Binder}{Prokopec
  et~al\mbox{.}}{2019b}]%
        {pldi-prokopec-19}
\bibfield{author}{\bibinfo{person}{Aleksandar Prokopec},
  \bibinfo{person}{Andrea Rosà}, \bibinfo{person}{David Leopoldseder},
  \bibinfo{person}{Gilles Duboscq}, \bibinfo{person}{Petr Tůma},
  \bibinfo{person}{Martin Studener}, \bibinfo{person}{Lubomír Bulej},
  \bibinfo{person}{Yudi Zheng}, \bibinfo{person}{Alex Villazón},
  \bibinfo{person}{Doug Simon}, \bibinfo{person}{Thomas Würthinger}, {and}
  \bibinfo{person}{Walter Binder}.} \bibinfo{year}{2019}\natexlab{b}.
\newblock \showarticletitle{Renaissance: Benchmarking Suite for Parallel
  Applications on the JVM}.
\newblock  (\bibinfo{year}{2019}).
\newblock


\bibitem[\protect\citeauthoryear{Ratanaworabhan, Livshits, and
  Zorn}{Ratanaworabhan et~al\mbox{.}}{2010}]%
        {Ratanaworabhan:2010:JCB:1863166.1863169}
\bibfield{author}{\bibinfo{person}{Paruj Ratanaworabhan},
  \bibinfo{person}{Benjamin Livshits}, {and} \bibinfo{person}{Benjamin~G.
  Zorn}.} \bibinfo{year}{2010}\natexlab{}.
\newblock \showarticletitle{{JSMeter: Comparing the Behavior of JavaScript
  Benchmarks with Real Web Applications}}. In
  \bibinfo{booktitle}{\emph{WebApps}}. \bibinfo{pages}{3--3}.
\newblock


\bibitem[\protect\citeauthoryear{Sch\"{o}rgenhumer, Hofer, Gnedt, and
  M\"{o}ssenb\"{o}ck}{Sch\"{o}rgenhumer et~al\mbox{.}}{2017}]%
        {Schorgenhumer:2017:ESL:3030207.3030234}
\bibfield{author}{\bibinfo{person}{Andreas Sch\"{o}rgenhumer},
  \bibinfo{person}{Peter Hofer}, \bibinfo{person}{David Gnedt}, {and}
  \bibinfo{person}{Hanspeter M\"{o}ssenb\"{o}ck}.}
  \bibinfo{year}{2017}\natexlab{}.
\newblock \showarticletitle{{Efficient Sampling-based Lock Contention Profiling
  for Java}}. In \bibinfo{booktitle}{\emph{ICPE}}. \bibinfo{pages}{331--334}.
\newblock


\bibitem[\protect\citeauthoryear{Sewe, Mezini, Sarimbekov, Ansaloni, Binder,
  Ricci, and Guyer}{Sewe et~al\mbox{.}}{2012}]%
        {Sewe:2012}
\bibfield{author}{\bibinfo{person}{Andreas Sewe}, \bibinfo{person}{Mira
  Mezini}, \bibinfo{person}{Aibek Sarimbekov}, \bibinfo{person}{Danilo
  Ansaloni}, \bibinfo{person}{Walter Binder}, \bibinfo{person}{Nathan Ricci},
  {and} \bibinfo{person}{Samuel~Z. Guyer}.} \bibinfo{year}{2012}\natexlab{}.
\newblock \showarticletitle{{new Scala() instanceof Java: A Comparison of the
  Memory Behaviour of Java and Scala Programs}}. In
  \bibinfo{booktitle}{\emph{ISMM}}. \bibinfo{pages}{97--108}.
\newblock


\bibitem[\protect\citeauthoryear{Sewe, Mezini, Sarimbekov, and Binder}{Sewe
  et~al\mbox{.}}{2011}]%
        {Sewe:2011:DCC:2048066.2048118}
\bibfield{author}{\bibinfo{person}{Andreas Sewe}, \bibinfo{person}{Mira
  Mezini}, \bibinfo{person}{Aibek Sarimbekov}, {and} \bibinfo{person}{Walter
  Binder}.} \bibinfo{year}{2011}\natexlab{}.
\newblock \showarticletitle{{Da Capo Con Scala: Design and Analysis of a Scala
  Benchmark Suite for the Java Virtual Machine}}. In
  \bibinfo{booktitle}{\emph{OOPSLA}}. \bibinfo{pages}{657--676}.
\newblock


\bibitem[\protect\citeauthoryear{Stadler, W\"{u}rthinger, and
  M\"{o}ssenb\"{o}ck}{Stadler et~al\mbox{.}}{2014}]%
        {Stadler:2014:PEA:2581122.2544157}
\bibfield{author}{\bibinfo{person}{Lukas Stadler}, \bibinfo{person}{Thomas
  W\"{u}rthinger}, {and} \bibinfo{person}{Hanspeter M\"{o}ssenb\"{o}ck}.}
  \bibinfo{year}{2014}\natexlab{}.
\newblock \showarticletitle{{Partial Escape Analysis and Scalar Replacement for
  Java}}. In \bibinfo{booktitle}{\emph{CGO}}.
  \bibinfo{pages}{165:165--165:174}.
\newblock


\bibitem[\protect\citeauthoryear{Thiessen and Lhot\'{a}k}{Thiessen and
  Lhot\'{a}k}{2017}]%
        {Thiessen:2017:CTP:3062341.3062359}
\bibfield{author}{\bibinfo{person}{Rei Thiessen} {and}
  \bibinfo{person}{Ond\v{r}ej Lhot\'{a}k}.} \bibinfo{year}{2017}\natexlab{}.
\newblock \showarticletitle{{Context Transformations for Pointer Analysis}}. In
  \bibinfo{booktitle}{\emph{PLDI}}. \bibinfo{pages}{263--277}.
\newblock


\bibitem[\protect\citeauthoryear{\v{S}ev\v{c}\'{\i}k}{\v{S}ev\v{c}\'{\i}k}{2011}]%
        {Sevcik:2011:SOS:1993498.1993534}
\bibfield{author}{\bibinfo{person}{Jaroslav \v{S}ev\v{c}\'{\i}k}.}
  \bibinfo{year}{2011}\natexlab{}.
\newblock \showarticletitle{Safe Optimisations for Shared-memory Concurrent
  Programs}. In \bibinfo{booktitle}{\emph{Proceedings of the 32Nd ACM SIGPLAN
  Conference on Programming Language Design and Implementation}}
  \emph{(\bibinfo{series}{PLDI '11})}. \bibinfo{publisher}{ACM},
  \bibinfo{address}{New York, NY, USA}, \bibinfo{pages}{306--316}.
\newblock
\showISBNx{978-1-4503-0663-8}
\urldef\tempurl%
\url{https://doi.org/10.1145/1993498.1993534}
\showDOI{\tempurl}


\bibitem[\protect\citeauthoryear{Weninger and M\"{o}ssenb\"{o}ck}{Weninger and
  M\"{o}ssenb\"{o}ck}{2018}]%
        {Weninger:2018:UCM:3184407.3184412}
\bibfield{author}{\bibinfo{person}{Markus Weninger} {and}
  \bibinfo{person}{Hanspeter M\"{o}ssenb\"{o}ck}.}
  \bibinfo{year}{2018}\natexlab{}.
\newblock \showarticletitle{{User-defined Classification and Multi-level
  Grouping of Objects in Memory Monitoring}}. In
  \bibinfo{booktitle}{\emph{ICPE}}. \bibinfo{pages}{115--126}.
\newblock


\bibitem[\protect\citeauthoryear{Wood, Cao, Bond, and Grossman}{Wood
  et~al\mbox{.}}{2017}]%
        {Wood:2017:IBD:3152284.3133893}
\bibfield{author}{\bibinfo{person}{Benjamin~P. Wood}, \bibinfo{person}{Man
  Cao}, \bibinfo{person}{Michael~D. Bond}, {and} \bibinfo{person}{Dan
  Grossman}.} \bibinfo{year}{2017}\natexlab{}.
\newblock \showarticletitle{{Instrumentation Bias for Dynamic Data Race
  Detection}}.
\newblock \bibinfo{journal}{\emph{Proc. ACM Program. Lang.}}
  \bibinfo{volume}{1}, \bibinfo{number}{OOPSLA} (\bibinfo{date}{Oct.}
  \bibinfo{year}{2017}), \bibinfo{pages}{69:1--69:31}.
\newblock


\bibitem[\protect\citeauthoryear{Zaharia, Chowdhury, Franklin, Shenker, and
  Stoica}{Zaharia et~al\mbox{.}}{2010}]%
        {Zaharia:2010:SCC:1863103.1863113}
\bibfield{author}{\bibinfo{person}{Matei Zaharia}, \bibinfo{person}{Mosharaf
  Chowdhury}, \bibinfo{person}{Michael~J. Franklin}, \bibinfo{person}{Scott
  Shenker}, {and} \bibinfo{person}{Ion Stoica}.}
  \bibinfo{year}{2010}\natexlab{}.
\newblock \showarticletitle{{Spark: Cluster Computing with Working Sets}}. In
  \bibinfo{booktitle}{\emph{HotCloud}}. \bibinfo{pages}{10--10}.
\newblock


\end{thebibliography}


\clearpage

\appendix

\end{document}